\documentclass[sigconf]{acmart}
\usepackage{makecell}
\usepackage{multirow}
\usepackage{subcaption}
\usepackage{float}
\usepackage{graphicx}
\usepackage{amsmath,amsfonts}
\usepackage{nicefrac}       
\usepackage{listings}

\usepackage{pifont}
\newcommand{\cmark}{\ding{51}}%
\AtBeginDocument{%
  }

\copyrightyear{2023}
\acmYear{2023}
\setcopyright{rightsretained}
\acmConference[MICRO '23]{56th Annual IEEE/ACM International Symposium on Microarchitecture}{October 28-November 1, 2023}{Toronto, ON, Canada}
\acmBooktitle{56th Annual IEEE/ACM International Symposium on Microarchitecture (MICRO '23), October 28-November 1, 2023, Toronto, ON, Canada}
\acmDOI{10.1145/3613424.3623797}
\acmISBN{979-8-4007-0329-4/23/10}

\newcommand{\rebuttal}[1]{#1}
\newcommand{\shepherding}[1]{#1}

\newcommand{\sys}{DOSA}

\newcommand{\lstprompt}{\$}
\newcommand\prompt[1]{\small\ttfamily\selectfont \lstprompt}

\usepackage{acmart-taps}
\usepackage{framed}

\begin{document}

\title[DOSA: Differentiable Model-Based One-Loop Search for DNN Accelerators]{\sys{}: \underline{D}ifferentiable Model-Based \\ \underline{O}ne-Loop \underline{S}earch for DNN \underline{A}ccelerators}


\author{Charles Hong}
\affiliation{%
  \institution{University of California, Berkeley}
  \city{Berkeley}
  \state{CA}
  \country{USA}
}
\email{charleshong@berkeley.edu}

\author{Qijing Huang}
\affiliation{%
  \institution{NVIDIA}
  \city{Santa Clara}
  \state{CA}
  \country{USA}
}
\email{jennyhuang@nvidia.com}

\author{Grace Dinh}
\affiliation{%
 \institution{University of California, Berkeley}
 \city{Berkeley}
 \state{CA}
 \country{USA}
 }
\email{dinh@berkeley.edu}

\author{Mahesh Subedar}
\affiliation{%
 \institution{Intel Labs}
 \city{Hillsboro}
 \state{OR}
 \country{USA}
 }
\email{mahesh.subedar@intel.com}

\author{Yakun Sophia Shao}
\affiliation{%
 \institution{University of California, Berkeley}
 \city{Berkeley}
 \state{CA}
 \country{USA}
 }
\email{ysshao@berkeley.edu}

\renewcommand{\shortauthors}{Hong, Huang, Dinh, Subedar, and Shao}

\begin{abstract}
In the hardware design space exploration process, it is critical to optimize both hardware parameters and algorithm-to-hardware mappings.
Previous work has largely approached this simultaneous optimization problem by separately exploring the hardware design space and the mapspace---both individually large and highly nonconvex spaces---independently. The resulting combinatorial explosion has created significant difficulties for optimizers. 

In this paper, we introduce \sys{}, which consists of differentiable performance models and a gradient descent-based optimization technique to simultaneously explore both spaces and identify high-performing design points. Experimental results demonstrate that \sys{} outperforms random search and Bayesian optimization by 2.80$\times$ and 12.59$\times$, respectively, in improving DNN model energy-delay product, given a similar number of samples. We also demonstrate the modularity and flexibility of \sys{} by augmenting our analytical model with a learned model, allowing us to optimize buffer sizes and mappings of a real DNN accelerator and attain a 1.82$\times$ improvement in energy-delay product. 


\end{abstract}





\maketitle

\begin{figure}[t] 
\vspace{10pt}
    \centering
    \begin{minipage}{0.55\columnwidth}
        \centering
        \includegraphics[width=0.75\columnwidth]{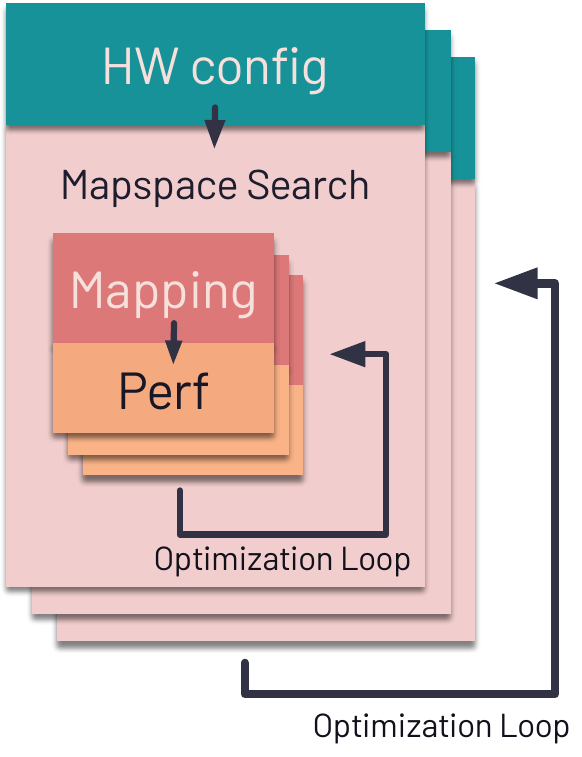} 
    \end{minipage}\hfill
    \begin{minipage}{0.45\columnwidth}
        \centering
        \includegraphics[width=0.75\columnwidth]{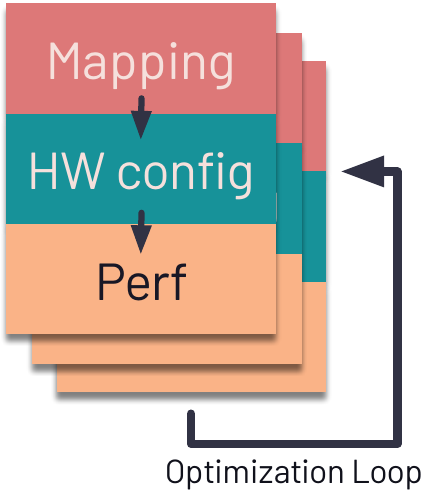} 
    \end{minipage}
\caption{Hardware-first, two-loop (left) and mapping-first, one-loop (right) DSE approaches.}
\label{fig:hw-first-mapping-first}
\end{figure}

\section{Introduction}
Deep neural network (DNN) accelerators~\cite{nvidia_a100, tpu-isca2016, lauterbach2021path, aws-inferentia} have become a critical driving force for the recent breakthroughs~\cite{alexnet, goodfellow2014generative, resnet, transformer, chatgpt} in artificial intelligence. 
To develop efficient DNN accelerators in a fast and cost-effective manner, automated design space exploration (DSE) has emerged as a powerful technique. 

The hardware DSE flow~\cite{venkatesan2019magnet, parashar2019timeloop, kwon2020maestro} involves optimizing over two search spaces: the \emph{hardware design space}, which describes hardware design parameters such as interconnect topology and buffer and systolic array sizes, and the \emph{mapspace}, which describes how applications are executed on the target hardware and encompasses decisions such as tiling, dataflow, and spatio-temporal mapping.

For both the hardware design space and the mapspace, the goal is to optimize a performance metric, such as energy-delay product (EDP), subject to certain constraints. These include design budgets, such as bounds on the area or power consumption, as well as constraints ensuring that the selected mapping can be executed on the selected hardware configuration (e.g. that the hardware buffers are sufficiently large to contain the tiles). As these constraints encompass both the mapspace and the hardware design space, the two spaces must be simultaneously optimized over; techniques solely tackling hardware search \cite{kumar2021prime,yazdanbakhsh2021apollo} or mapping optimizations~\cite{paszke2019pytorch, abadi2016tensorflow, tensorrt, chen2015mxnet, chen2018tvm, sabne2020xla, kjolstad2017tensor} are insufficient to achieve the optimal hardware design in DSE.

Both the hardware design and mapping spaces are vast, high-dimensional, and comprised of both categorical and discrete variables. Furthermore, evaluating the performance of a hardware configuration and a mapping can be computationally expensive. The size of the combined optimization space and the cost of evaluating points in it pose formidable challenges to DSE algorithms.

Much prior work~\cite{venkatesan2019magnet, zhang2022full, huang2022learning, sakhuja2023spotlight, xiao2021hasco, lin2021naas} has approached this problem using \emph{hardware-first search}. These methods directly search over the space of possible hardware configurations. The performance of each hardware configuration is calculated by first constraining the mapspace to mappings that are compatible with the hardware configuration, then optimizing over the constrained (highly discontinuous) mapspace. In most cases, the mapspace optimization is done iteratively, rendering this process a \emph{two-loop approach} iterating over both the hardware space and mapspace. As a result, these approaches must contend with a combinatorial explosion of possible configurations.

Alternatively, \emph{mapping-first} approaches, as proposed in~\cite{kao2022digamma, interstellar-asplos2020} and illustrated in Figure~\ref{fig:hw-first-mapping-first}, optimize primarily over the mapspace. For each mapping, optimizing over the hardware design space is a straightforward process consisting of finding the minimal hardware configuration capable of supporting the mapping. As a result, the loop for hardware search is eliminated, allowing the entire DSE process to be encapsulated in a single loop. Furthermore, the lack of hardware resource constraints also significantly simplifies the mapspace search problem.

Despite these advantages, mapping-first approaches must still contend with the size of the mapspace and the nonconvexity of the performance over this space. Prior works have either directly applied black-box optimization methods~\cite{kao2022digamma, shi2020using}, which rely on a large number of (often expensive to collect) samples, or pruned the search space using architecture-specific heuristics constructed by hand based on observations of a limited subset of the DSE search space~\cite{interstellar-asplos2020}.

Reducing the sample complexity of DSE while still allowing for a systematic exploration of the entire space requires leveraging domain knowledge---for instance, a generalizable performance model like Timeloop~\cite{parashar2019timeloop, accelergy} (a popular analytical model for DNN accelerators), that is not reliant on an expensive training process. This paper follows this approach, using performance models as an optimization target for mapping-first search. Specifically:
\begin{itemize}
\item We build closed-form \emph{differentiable} and interpretable performance models for latency and energy on DNN accelerators. Our models are as precise as the state-of-the-art program-based analytical models, while also being amenable to white-box optimization techniques such as gradient descent.
\item We then introduce \sys{}\footnote{Code open-sourced at \url{https://github.com/ucb-bar/dosa}.\\Artifacts available: see Appendix~\ref{app:artifact}.}, a mapping-first one-loop DSE flow that uses gradient descent to find the most efficient hardware parameters and mappings to target multi-layer DNNs; to the best of our knowledge, this is the first work to use a mapping-first strategy to simultaneously perform DSE for multiple layers of a neural network. \sys{} converges at least 40\% faster than state-of-the-art DSE approaches.
\item We take a step beyond DSE for architectural model by introducing a DNN model to predict the variation between analytical model and real hardware accelerator performance, and using it to augment our differentiable model for real hardware DSE.
\item We benchmark our results on the Gemmini accelerator, showing a 1.82$\times$ EDP improvement over hand-designed configurations.
\end{itemize}

\section{Background}
\begin{table}[t]
\centering
\small\tabcolsep4pt\begin{tabular}{|l|>{\centering}p{60pt}|cc|}
\hline
\textbf{} &
  \multicolumn{1}{c|}{\textbf{Name}} &
  \multicolumn{1}{c|}{\textbf{\begin{tabular}[c]{@{}c@{}}Mapspace \\ Search\end{tabular}}} &
  \textbf{\begin{tabular}[c]{@{}c@{}}Hardware  \\ Search\end{tabular}} \\ \hline
\multirow{7}{*}{\textbf{\begin{tabular}[c]{@{}l@{}}Two-loop \\ Searchers\end{tabular}}} &
  Spotlight~\cite{sakhuja2023spotlight} &
  \multicolumn{1}{c|}{BB-BO} &
  BB-BO \\ \cline{2-4} 
 & VAESA~\cite{huang2022learning}     & \multicolumn{1}{c|}{ILP~\cite{cosa2021huang}}                & VAE+BB-BO/GD \\ \cline{2-4} 
 & FAST~\cite{zhang2022full}          & \multicolumn{1}{c|}{BB-LCS~\cite{karro2017black}+ILP} & BB-LCS    \\ \cline{2-4} 
 & HASCO~\cite{xiao2021hasco}         & \multicolumn{1}{c|}{RL}                                      & BB-BO        \\ \cline{2-4} 
 & NAAS~\cite{lin2021naas}            & \multicolumn{1}{c|}{BB-ES}                                   & BB-ES        \\ \cline{2-4} 
 & MAGNet~\cite{venkatesan2019magnet} & \multicolumn{1}{c|}{Heuristics}                              & BB-BO        \\ \hline

\multirow{4}{*}{\textbf{\begin{tabular}[c]{@{}l@{}}One-loop \\ Searchers\end{tabular}}} 
& DiGamma~\cite{kao2022digamma} & \multicolumn{2}{c|}{BB-GA} \\ \cline{2-4}
& Interstellar~\cite{interstellar-asplos2020} & \multicolumn{2}{c|}{ Heuristics} \\ \cline{2-4}
& \textbf{Our work:}\newline\textbf{\sys{}}    & \multicolumn{2}{c|}{\multirow{2}{*}{\textbf{GD}}} \\
\hline
\end{tabular}
\caption{State-of-the-art Accelerator DSE Methods.\label{table:related_work}}\vspace{-12pt}
\end{table}

Hardware design space exploration (DSE) is a time-consuming and costly process that involves the exploration of various hardware design parameters and software mappings to optimize the target application performance. 
This process typically includes two key optimizations: the mapping search, which aims to find high-performance mappings that effectively utilize hardware resources, and the hardware  search, which aims to achieve multi-objective design goals, such as minimizing the energy-delay product (EDP) or the area-delay product. 
To address the mapping complexity for DNNs, many DNN compilers~\cite{paszke2019pytorch, abadi2016tensorflow, tensorrt, chen2015mxnet, chen2018tvm, sabne2020xla, kjolstad2017tensor} and accelerator-aware mapping techniques~\cite{parashar2019timeloop, interstellar-asplos2020, gamma-iccad2020, hegde2021mind, cosa2021huang, li2021analytical} have been developed. 
In addition, there has been extensive research in the area of hardware parameter search~\cite{kumar2021prime,yazdanbakhsh2021apollo}.



\subsection{Co-exploration Frameworks}\label{sec:bg-coexploration}
In recent years, there has been a growing body of research focused on tackling the compounding search space of mapping and hardware designs with the goal of achieving higher hardware efficiency and lower development costs.
 
\subsubsection{Two-Loop Searchers}
As shown in Table~\ref{table:related_work}, most prior work~\cite{sakhuja2023spotlight, huang2022learning, zhang2022full, xiao2021hasco, lin2021naas, venkatesan2019magnet, interstellar-asplos2020} treats the mapping and hardware co-search as a two-loop process and applies a combination of various optimization techniques to address each search space independently. 
The two-loop process starts by sampling a hardware design point from the hardware search space and then searching for high-performance mappings for that particular hardware design point in the inner loop. The best mapping obtained is used for generating the hardware performance feedback for the outer loop hardware optimization.


Optimization techniques used in the two-loop process can be broadly categorized into three types: heuristics, black-box optimization (BB), and white-box optimization.
Heuristics involve using domain-specific knowledge and experience to guide the search process and reduce the size of design space. In contrast, BB relies on sampling and machine learning techniques to leverage the characteristics of the problem derived from sampled data in order to find the optimal solution. Popular BB algorithms include genetic algorithms (BB-GA), reinforcement learning (RL), Bayesian optimization (BB-BO), Linear Combination Swarm (BB-LCS), and evolutionary strategy (BB-ES).

In white-box optimization, the relationship between the optimization variables and the objectives is known and captured in mathematical models. 
Numerical optimization techniques like linear programming (LP) and mixed-integer programming (MIP) can be used if the relationship can be expressed in specific frameworks. 
Gradient descent (GD) techniques can be applied if the relationship can be expressed in a differentiable expression.  
Compared to black-box optimization, white-box optimization is generally more efficient as it can exploit the known objective model to guide the optimization
process, resulting in faster convergence. 
However, it requires the objective model to be known and accurately specified.

While independently applying optimization techniques to the mapspace and hardware space can be effective, the two-loop searchers can be susceptible to combinatorial explosion, as the vast search space multiplies the number of potential options for mappings and hardware parameters together. 

\subsubsection{Single-Loop Searchers}

To reduce the size of the compounding search space, one-loop searchers, such as DiGamma~\cite{kao2022digamma} and Interstellar~\cite{interstellar-asplos2020} have been proposed. Single-loop search tackles the co-search problem from a mapping-first approach that infers the minimal hardware requirement from hardware-agnostic high-performance mappings found in single-loop mapping search. In such approaches, the hardware DSE space is similar in size to the mapping space. However, DiGamma employs BB-GA which treats the mapping performance as a black-box and needs to evaluate many unique hardware and mapping configurations iteratively to achieve a good mapping and hardware design. 
Interstellar, on the other hand, only explores a limited space of pre-selected mappings and as a result only a limited space of hardware design is evaluated.

Unlike previous one-loop approaches that rely on black-box optimizations or heuristics, \sys{} takes a novel approach by formulating the analytical performance and energy model in ~\cite{parashar2019timeloop} as a differentiable white-box model. \sys{} uses gradient descent to optimize the mapping variables in the direction of steepest descent of the EDP objective function on the mathematical model. This allows \sys{} to explore a comprehensive set of mappings and efficiently generate high-quality hardware and mapping configurations without the need for sampling from simulators extensively.

\subsection{Performance Modeling}\label{sec:bg-modeling}
Performance models are crucial to the DSE process, as they offer quick feedback on performance and energy consumption for different hardware and mappings. They provide valuable insights into how hardware designs perform in real-world scenarios without requiring real hardware prototypes or implementations. They can also reduce the high sampling cost of time-consuming and computationally expensive simulation and emulation for existing hardware designs.
In this paper, we illustrate how a well-designed performance model can be used to accelerate the DSE process.  

Depending on how they are developed, performance models can be categorized as either analytical models or data-driven models. For DNN accelerators, architects have developed domain-specific analytical models in the form of mathematical equations~\cite{cosa2021huang} or iterative programs~\cite{parashar2019timeloop, kwon2020maestro, lu2021tenet, mei2021zigzag,  wu2022sparseloop} to quickly assess tradeoffs for various hardware designs. 
These models leverage workload characteristics (e.g., known iteration space bounds and statically analyzable data access patterns), and hardware characteristics (e.g. roofline model~\cite{williams2009roofline}) to perform the estimation. Data-driven models~\cite{chen2018tvm, hegde2021mind, kaufman2021learned}, on the other hand, use statistical techniques to fit a machine learning (ML) model to performance data collected over time. 

Different models offer different levels of fidelity and compatibility with DSE optimization algorithms. Iterative programs are often used with BB algorithms as the relationship between inputs and predictions is not directly known to the optimization algorithms, which rely on sampling to recapture this relationship. 
Analytical models expressed in mathematical equations can be used directly as objectives in optimization, but the existing formulations in linear or quadratic programs~\cite{cosa2021huang} tend to be limited in expressiveness for complex hardware systems and can result in low accuracy.
ML models can be integrated with various optimization techniques easily, but they typically need a large amount of training data to provide accurate prediction and generalize to new workloads and architectures. 

This work aims to improve upon existing performance models by introducing a differentiable performance model that is highly accurate, generalizable, and amenable to various efficient white-box optimization algorithms, such as GD. Our approach involves decoding the mathematical relationships from Timeloop~\cite{parashar2019timeloop}, an accurate iterative program-based analytical model for DNN accelerators, and converting part of the iterative program into differentiable mathematical models. We show that, by making the objectives differentiable with respect to the design parameters, our approach achieves high sample efficiency in DSE.

While existing analytical models for DNN accelerators are trusted by architects, they may not capture all the interactions between the accelerator and the rest of the hardware systems in real-world deployment. To address this, this work also introduces an ML model that predicts the differences between the analytical model and the actual hardware, thereby improving the effectiveness of DSE in a real-world setting.



\section{\sys{} Overview}

\begin{figure}[t]
    \centering
    \includegraphics[width=0.99\columnwidth]{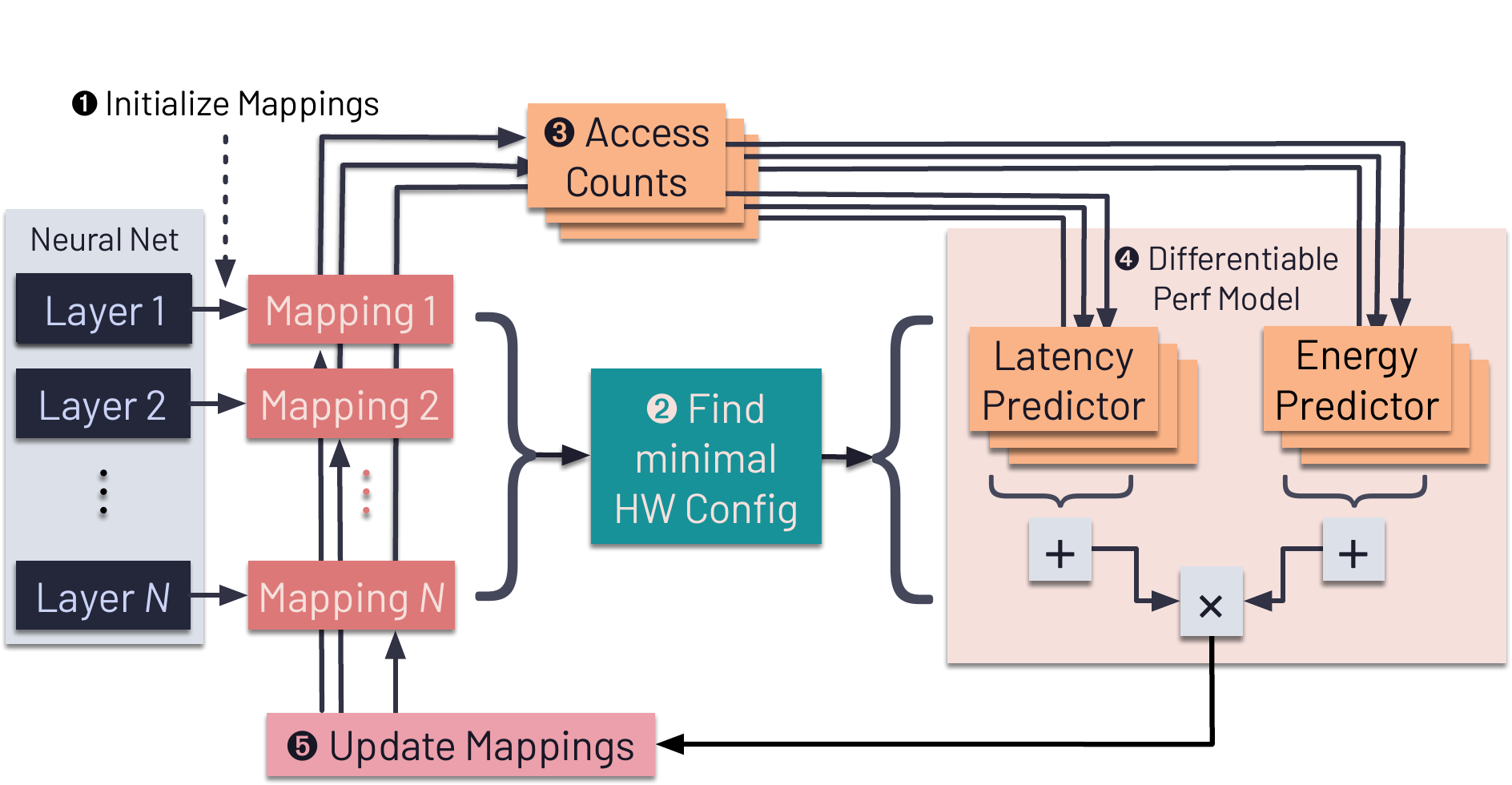}
    \caption{An architecture diagram of \sys{}.}\label{fig:overview}
\end{figure}

This paper presents \sys{}, a one-loop differentiate-model-based DSE framework to optimize the mappings and hardware simultaneously for target DNN models. 
\sys{} captures key relations between DNN mapping factors and performance objectives in a differentiable analytical model. 
In addition, \sys{} introduces a data-driven DNN model to capture the performance variations between analytical model and real hardware.
By applying white-box optimization to the model and calculating the hardware parameters using minimal parameterization,
\sys{} achieves high-performance accelerator design and mapping while significantly reducing the time and costs associated with DNN accelerator DSE. 

\subsection{Problem Setup}

\subsubsection{Target Workloads} \sys{} targets accelerator DSE for complete DNN models, which comprise both matrix multiplication and convolution layers. \rebuttal{To express these layers, we use seven dimensions: $R$ (weight height), $S$ (weight width), $P$ (output activation height), $Q$ (output activation width), $C$ (input channels), $K$ (output channels), and $N$ (batch size). These dimensions describe the size of the weight ($W$), input ($I$), and output ($O$) tensors.}
\rebuttal{We assume the activation functions are fused with the matrix multiplication and convolution layers.}

\subsubsection{Variables and Objectives}\label{sec:mapping} In our mapping-first search, we focus on the following three layerwise mapping decisions:
\begin{enumerate}
    \item Spatial loop tiling, which defines which loops are mapped to parallel spatial resources (such as processing elements in a systolic array), and the \rebuttal{iteration} bounds of these loops,
    \item Temporal loop tiling, \rebuttal{which specifies the loop iteration bounds grouped together to form a block at each memory level.}
    \item Loop ordering, which defines the order in which dimensions are accessed at a given memory level.
\end{enumerate}

We utilize the spatial and temporal tiling factors, denoted as {\small$\vec{f}$}, as input optimization variables in our approach. Specifically, for dimension $d$ at memory level $i$, $f_{S,i,d}$ and $f_{T,i,d}$ represent the spatial and temporal tiling factors, respectively. Using {\small$\vec{f}$}, we construct \sys{}'s objective function, which serves as the analytical performance model predicting energy-delay product (EDP) of the DNN, as detailed in Section~\ref{sec:methodology}. To optimize performance, gradient descent is employed to differentiate the objective with respect to the tiling variables {\small$\vec{f}$}. The optimization details are elaborated on in Section~\ref{sec:optimization}.  Note that there are constraints imposed on the variables to ensure that for each dimension, the product of the spatial and temporal tiling factors at all memory levels is equal to the total problem size.

\subsection{Toolflow}
Figure~\ref{fig:overview} provides an overview of how \sys{} simultaneously optimizes mappings and hardware for a given workload consisting of a set of layers. The following are the detailed steps involved in this process:

\begin{enumerate}
\item  Generate performant mappings using CoSA~\cite{cosa2021huang} for a set of target DNN layers, targeting a randomly selected valid hardware design.
\item Compute the hardware resource requirements of the layerwise mappings and convert them to a minimal hardware parameterization.
\item Given the mappings represented in {\small$\vec{f}$}, use the differentiable model in \sys{} to calculate \rebuttal{the number of arithmetic operations and} the number of accesses made by each mapping to each memory level in the accelerator.
\item Combine \rebuttal{arithmetic operation and memory access counts} with previously calculated hardware parameters to generate roofline-based latency predictions and event-based energy predictions for each layer's mapping. Then, each mapping's latency and energy prediction is combined to produce a single EDP value.
\item Use gradient descent to update all mappings in parallel.
\item Repeat from Step 2.
\end{enumerate}

\begin{figure}[t]
    \centering
    \includegraphics[trim={0.5cm 8.5cm 1.5cm 0.5cm}, width=0.95\columnwidth]{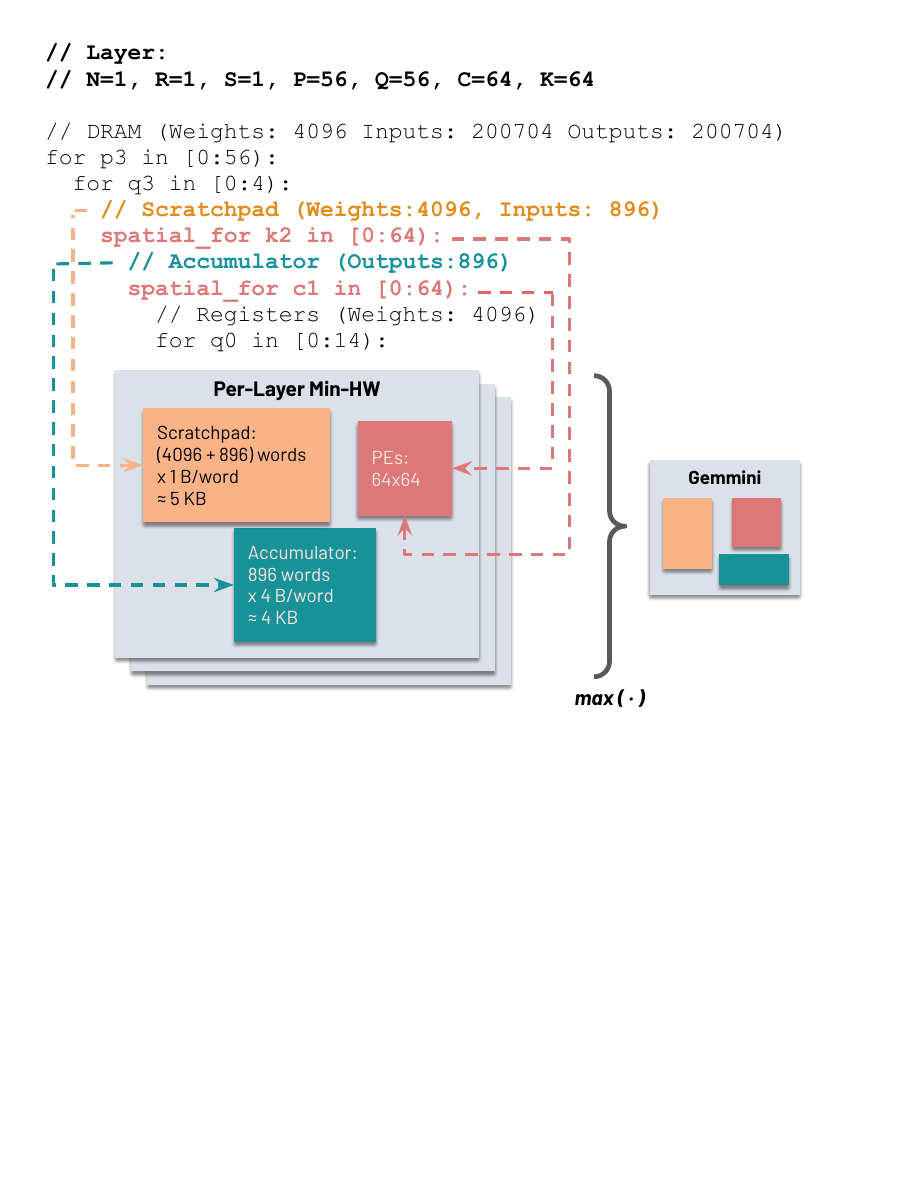}
    \caption{
     Mapping to hardware parameters conversion in \sys{}. The final hardware configuration is selected by taking the max across mappings for each hardware parameter.}\label{fig:min-hw}
\end{figure} 


\begin{table*}[ht]
    \centering
    \begin{tabular}{cccc}
    \toprule
    Arch. Component & Memory Level & Bandwidth (words/cycle) & Energy Per Access (EPA, in uJ)~\cite{cacti} \\
    \midrule
        PE & & & 0.561 \\
        Registers & $0$ & $2C_{PE}$ & 0.487 \\
        Accumulator & $1$ & $2\sqrt{C_{PE}}$ & $1.94 + 0.1005 \times \frac{C_1}{\sqrt{C_{PE}}}$ \\
        Scratchpad & $2$ & $2\sqrt{C_{PE}}$ & $0.49 + 0.025 \times C_2$ \\
        DRAM & $3$ & 8 & 100 \\
        \bottomrule
    \end{tabular}
    \caption{Details of the accelerator under study. 
    \rebuttal{$C_{PE}$ is the total number of PEs and $C_{i}$ is the capacity of memory level  $i$.}}\label{table:gemmini}
        \vspace{-12pt}
\end{table*}

\section{\sys{} Differentiable Model}\label{sec:methodology}

The differentiable model of \sys{} is motivated by the following observations:

\begin{itemize}
    \item \rebuttal{As discussed in Section~\ref{sec:bg-coexploration}, inferring hardware parameters from mappings flattens the hardware-mapping co-search space, and allows for mapping-first, one-loop search.}
    \item An effective mapping-first searcher should co-optimize mapping variables of all layers in the target DNN. This high-dimensional problem requires the application of an efficient optimization method such as gradient descent.
    \item A differentiable performance model can facilitate DSE with gradient descent, and constructing the model analytically is preferable as it ensures the accuracy, interpretability, and generalizability of the model.
\end{itemize}

Given the absence of a differentiable, analytical model for DNN accelerators in current literature, we present our approach for constructing such a model that achieves accuracy on par with Timeloop in our problem space. 
To account for performance variations in real hardware that are difficult to capture and express in analytical models, we in addition trained a differentiable DNN model to further improve the accuracy of the performance model. 



\subsection{Computing Hardware Resource Requirements}
\label{sec:resource_req}
We target the open-source DNN accelerator Gemmini~\cite{gemmini}, whose most notable architectural components are 1) a systolic array of processing elements (PEs), 2) accumulator SRAM, 3) scratchpad SRAM, and 4) DRAM. Specifically, we target the weight-stationary (WS) configuration of Gemmini. 
The buffer levels are enumerated $1, 2, ...$, where level $1$ represents the buffer memory level. 
Memory level $0$ represents the per-PE registers in the systolic array. 
The architectural components of Gemmini are further detailed in \tableautorefname{}~\ref{table:gemmini}. As depicted in Figure~\ref{fig:min-hw}, the capacity requirements at each level are first computed. Then, we take a parameter-wise max to generate a design that will support all current mappings.

\subsubsection{Notation}
In our notation, we use $i$ to index memory levels and $d$ to index problem dimensions for the spatial and temporal tiling factors $f_{S,i,d}$ and $f_{T,i,d}$, as listed in Table~\ref{tab:notation}. The spatial or temporal factor is indexed using $k$ in the subsequent section. Additionally, we use $t$ to index each data tensor.
\begin{table}[H]
\centering
\begin{tabular}{|l|l|}
\hline
$i$ & memory level index                 \\ \hline
$d$ & problem dimension index            \\ \hline
$k$ & spatial / temporal index           \\ \hline
$t$ & data tensor index                   \\ \hline
\end{tabular}
  \caption{Notation.}
  \label{tab:notation} \vspace{-16pt}
\end{table}

We define the following sets to express the consideration of problem dimensions for calculating the size of data tensors in DNN computations:
\begin{eqnarray*}
D & = & \{R,S,P,Q,C,K,N\}\\
D_{W} & = & \{R,S,C,K\}\\
D_{I} & = & \{R,S,P,Q,C,N\}\\
D_{0} & = & \{P,Q,K,N\}\\
\end{eqnarray*}
The set $D$ contains all the problem dimensions, while $D_W$, $D_I$, and $D_O$ are subsets of $D$ that contain the problem dimensions used to calculate the data tensor size and the minimal hardware requirements for weights, inputs, and outputs, respectively. 
\begin{equation*}
M  = \{0,1,2,3\}
\end{equation*}
$M$ is a set of indices that represents the memory levels available for storing intermediate tensors during the computation.
To keep track of which tensors are stored at each memory level, we define a matrix $B$ as shown in Table~\ref{tab:constant-matrix}. The entries of $B$ indicate whether a tensor with a certain problem dimension is stored at a certain memory level. 

\begin{table}[h]
\small
    \centering
    {\renewcommand{\arraystretch}{1.1}
    \begin{tabular}{c|c|c|c|c|c|}
    \multicolumn{2}{c|}{\multirow{2}{*}{ }} & \multicolumn{3}{c|}{ Tensor } \\
    \multicolumn{2}{c|}{} & W & I & O \\
    \hline 
    Registers & 0 & \cmark & & \\
    \hline
    Accumulator & 1 & & & \cmark \\
    \hline
    Scratchpad & 2 & \cmark & \cmark & \\
    \hline
    DRAM & 3 & \cmark & \cmark & \cmark \\
    \hline
    \end{tabular}}
  \caption{Constant binary matrix $B$, which encodes the data tensor(s) stored at each level of the memory hierarchy, for the accelerator under study.}
  \label{tab:constant-matrix} \vspace{-12pt}
\end{table}



\subsubsection{PE Capacity Requirements}

Gemmini supports only square arrays of processing elements. In its WS (weight stationary) configuration, it can parallelize the input channel ($C$ dimension) and output channel ($K$ dimension), each along one side of the array.  Hence, we need to configure a square PE array that is large enough to accommodate the square of the larger of these two spatial factors. The total number of processing elements in the systolic array is denoted by $C_{PE}$.
\begin{equation}
C_{PE} = max(f_{S,1,C}, f_{S,2,K})^2
\end{equation}

\subsubsection{Buffer Capacity Requirements}

Buffer capacities required at a given level $i$ for each tensor are computed by multiplying the related factors $f_{k,j,d}$ together. 

\begin{equation}
C_{i,W} = \prod_{(k,j,d)\in\{S,T\}\times\{i-1,i-2,...,0\}\times D_{W}}{f_{k,j,d}}\\
\end{equation}

\begin{equation}
\begin{gathered}
\text{Inner}(i,d) = \prod_{(k,j)\in\{S,T\}\times\{i-1,i-2,...,0\}}{f_{k,j,d}}\\
C_{i,I} = \left(\prod_{(k,j,d)\in\{S,T\}\times\{i-1,i-2,...,0\}\times\{C,N\}}{f_{k,j,d}}\right)\\
 \times\left(Pstride\times(\text{Inner}(i,P)-1)+\text{Inner}(i,R)\right)\\
 \times\left(Qstride\times(\text{Inner}(i,Q)-1)+\text{Inner}(i,S)\right)\\
\end{gathered}
\end{equation}

\begin{equation}
C_{i,O}  = \prod_{(k,j,d)\in\{S,T\}\times M\times D_{O}}{f_{k,j,d}}
\end{equation}

$C_{i,t}$ represents the number of words of tensor $t$ that memory level $i$ must be able to hold.
Note that to calculate the size required for inputs $C_{i,I}$, we first need to calculate the input activation dimensions using the stride and the output and weight dimensions factors (P,Q,R,S).

The total buffer capacity requirement at level $i$ is the sum of $C_{i,t}$ for each tensor $t$ that is stored at that level:
\begin{equation}
C_i = \sum_{t \in \{W,I,O\}}{B_{i,t}C_{i,t}}
\end{equation}
 
\subsection{Traffic Estimation}\label{sec:traf_est}
To capture the performance of the accelerator, we utilize differentiable non-convex functions to model the data movement at each buffer level.
We use the following terminologies to refer to different types of data transfer:
\begin{itemize}
    \item Writes - backing store memory to current memory
    \item Updates - faster memory or MAC to current memory
    \item Reads - current memory to faster memory or MAC
\end{itemize}


\subsubsection{Writes}
\rebuttal{The number of writes to a given memory level $i$ in Gemmini attributable to the tensor $t$ is given by multiplying the tensor size $C_{i,t}$ at level $i$ with all tiling factors outer to the innermost relevant loop, that is outer to level $i$.}
\begin{eqnarray}
\rebuttal{
    \text{Writes}_{t}(i) = C_{i,t}\prod_{\substack{(k,j,d) \in {S,T} \times{\{i+1,i+2,...,M\}} \\ \times D \text{ outer to } D_{t} > 1}} f_{k,j,d}
    }
\end{eqnarray}
\rebuttal{
For instance, to calculate the writes to weights, we can multiply the weight tensor size at memory level $i$, denoted as $C_{i,W}$, with all the tiling factors that are outer to the innermost loops R, S, C, or K given the loop order of all dimensions.
This calculation is performed similarly for the outputs. For inputs, as we did for the capacity calculation, we need to first compute the input factors by considering the stride and padding.
} 
The total tensor traffic at a given level is computed by summing weight, input, and output traffic.

\subsubsection{Updates}
Once write counts are computed, update traffic can be computed more easily.
Only outputs and partial sums incur updates to the current memory level $i$ from the inner memory or MAC. 
For every MAC operation, it will incur an output or partial sum update to the innermost memory level that stores the outputs. Therefore the number of updates to the innermost memory level is equal to the total number of MACs, which is defined as follows: 
\begin{equation}
    \text{MACs} =   \prod_{(k,j,d)\in\{S,T\}\times M\times D}{f_{k,j,d}}
\end{equation}
In the Gemmini architecture, as indicated in Table~\ref{tab:constant-matrix}, the innermost level corresponds to the accumulator at memory level 1. 
In the outer levels, the number of updates is equal to the number of writes \shepherding{to the next inner level that holds outputs,} as each time a partial sum is loaded, it undergoes addition and is subsequently stored back as an update. 
Note that the accumulation can also happen in the spatial network which will not result in an update to the memory. The overall reduction to the updates $F_{S,O}(i)$ can be determined by multiplying the spatial factors that are not related to the outputs: 
\begin{equation}
    F_{S,O}(i) = \prod_{d \in \{D-D_{O}\}}{f_{S,i,d}}
\end{equation}

Combining all these factors, the total updates at memory level $i$ can be expressed as:
\begin{equation}
\begin{gathered}
    \text{Updates}_{O}(i)  =  \begin{cases}
\frac{\text{MACs}}{F_{S,O}(i)} & i = \text{innermost output level} \\
\frac{\text{Writes}_{O}(\shepherding{i-1})}{F_{S,O}(i)} & i > \text{innermost output level} \\
\end{cases}\\
\text{Updates}(i) =  B_{i,O}\text{Updates}_{O}(i)
\end{gathered}
\end{equation}
where the MACs and Writes(i) are discounted by the factors that are spatially accumulated in the network. 

\subsubsection{Reads}
Similarly, when it comes to read operations, in the case of the innermost input buffer that holds inputs, the total number of reads is equal to the total number of MACs. This is because we need to load an input for each MAC calculation. For the outer levels, all the reads from the current level are transferred to the inner level as writes. In the presence of a broadcast spatial network at a certain level, if there are factors $F_{S,t}(i)$ that are irrelevant to the tensor, the same read operation will be broadcasted to different children, eliminating the need for multiple reads:

\begin{equation}
    F_{S,t}(i) = \prod_{d \in \{D-D_{t}\}}{f_{S,i,d}}
\end{equation}
Putting it all together, we have number of reads defined as: 
\begin{equation}
\begin{gathered}
\text{Reads}_{t}(i)  =  \begin{cases}
\frac{\text{MACs}}{F_{S,t}(i)} & i=\text{innermost tensor level} \\
\frac{\text{Writes}_{t}(i-1)}{F_{S,t}(i)} & i>\text{innermost tensor level} \\
\end{cases}\\
\text{Reads}(i)  =  \sum_{t\in\{W,I,O\}}{B_{i,t}\text{Reads}_{t}(i)}
\end{gathered}
\end{equation}

 \begin{figure*}[t]
  \centering
   \begin{subfigure}[b]{0.32\linewidth}
   \captionsetup{justification=centering}
    \includegraphics[width=\textwidth]{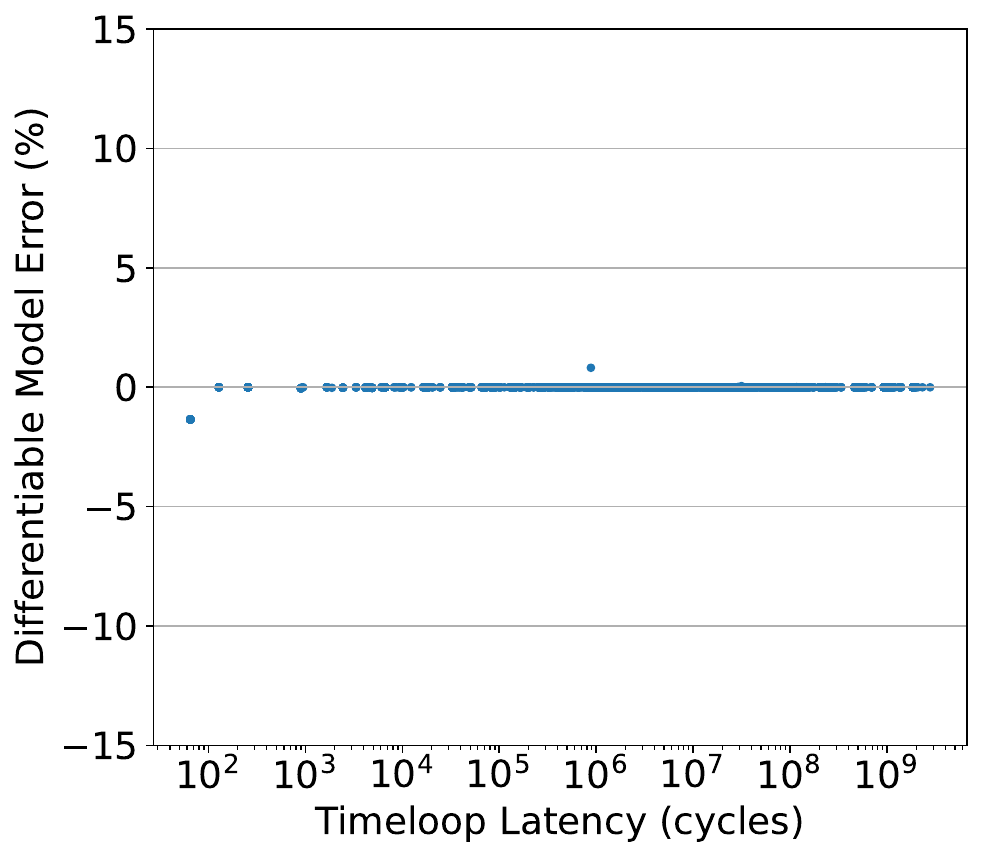}
        \caption{Latency: MAE=0.01\%}
  \end{subfigure}
  \begin{subfigure}[b]{0.32\linewidth}
  \captionsetup{justification=centering}
    \includegraphics[width=\textwidth]{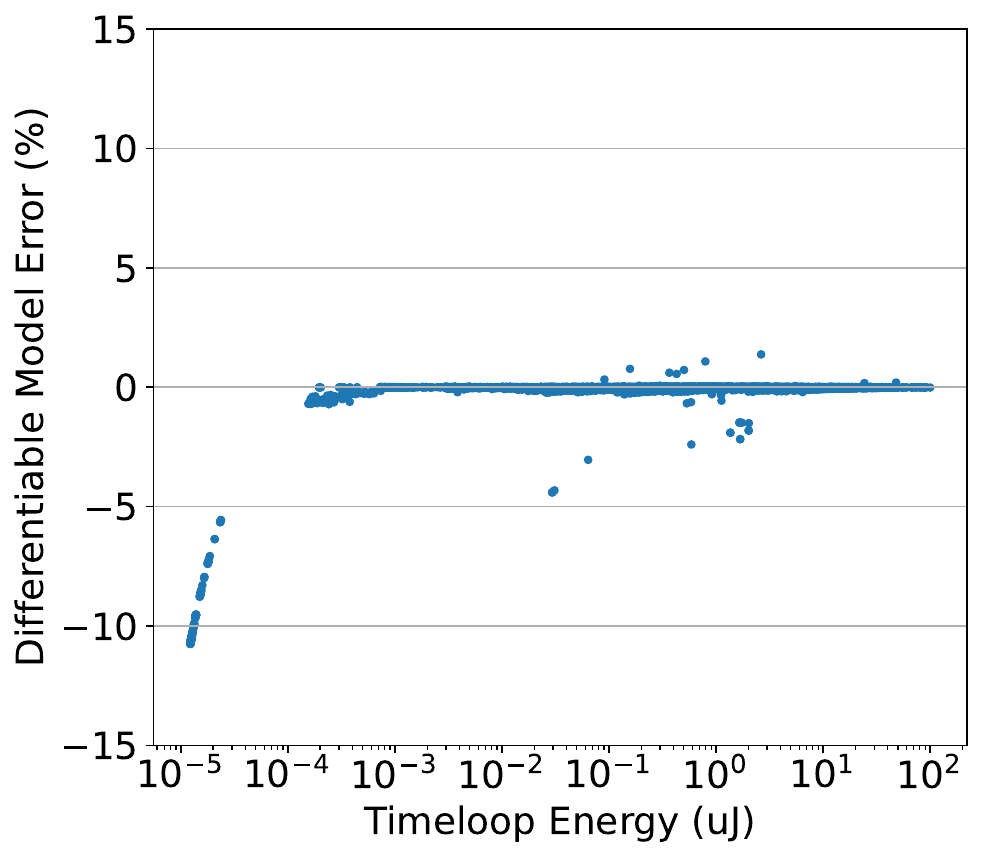}
        \caption{Energy: MAE=0.18\%}
  \end{subfigure}
  \begin{subfigure}[b]{0.32\linewidth}
    \includegraphics[width=\textwidth]{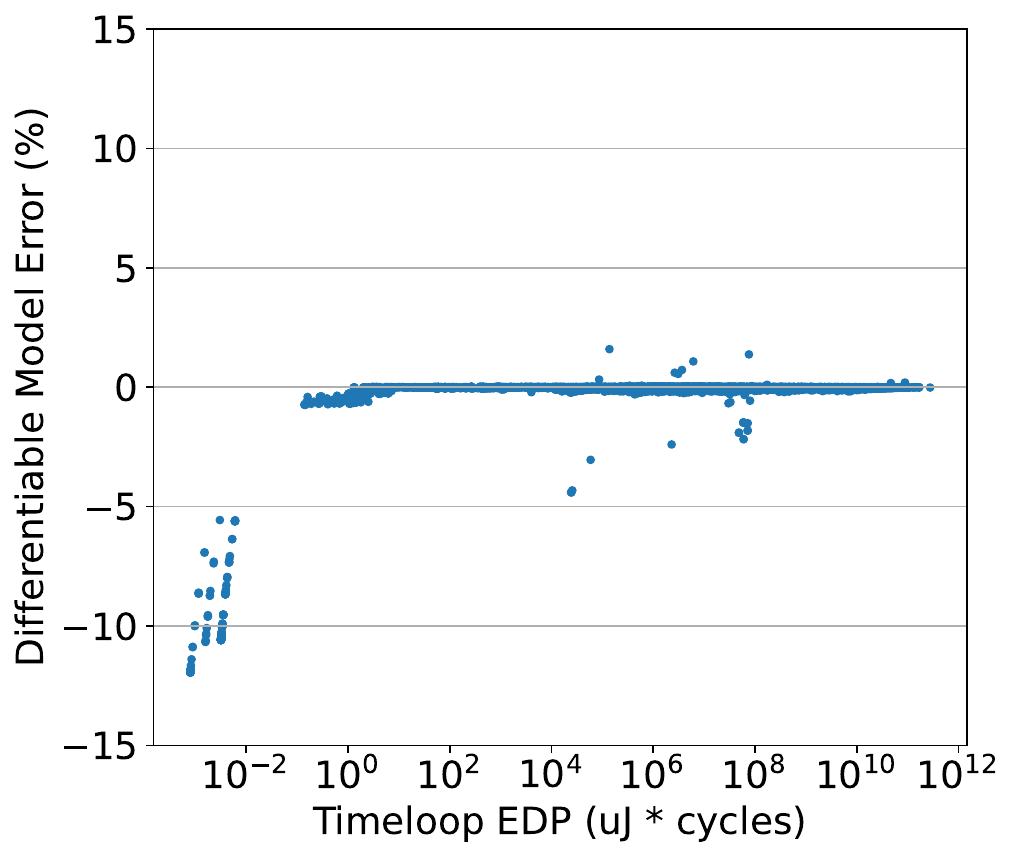}
        \caption{EDP: MAE=0.18\%}
  \end{subfigure}
  \caption{Error of \sys{} differentiable model prediction with respect to Timeloop for 100 random Gemmini configurations, 73 unique layers, 10,000 total mappings. MAE=Mean Absolute Error.}
  \label{fig:correlation}
\end{figure*}

\subsection{Latency Modeling}
\label{sec:latency-analytical}


We calculate the latency cycles required for compute by dividing the total number of multiply-accumulate (MAC) operations in a layer by the product of all spatial factors $f_{S,i,d}$ in a mapping (i.e., the number of parallel processing elements utilized).  
To compute memory access latency, we divide the total number of memory accesses by the memory bandwidth. We calculate the memory latency for each memory level $i$ utilized in Gemmini, including accumulator SRAM, scratchpad SRAM, and DRAM. We consider the maximum latency among all memory levels and the compute as the final latency since performance is limited either by memory or compute. The latency formulations are provided below:
%
\begin{equation}
\begin{gathered}
\text{Compute\_Latency}  =  \frac{\text{\# of MACs in Layer}}{\prod_{(i,d)\in M\times D}{f_{S,i,d}}}\\
\text{Accesses}(i)  =  \text{Reads}(i)+\text{Updates}(i)+\text{Writes}(i)\\
\text{Mem\_Latency}(i)  =  \frac{\text{Accesses}(i)}{\text{Bandwidth}(i)}\\
\text{Mem\_Latency} =  \max_{i\in M}(\text{Mem\_Latency}(i))\\
\text{Latency} =  \max(\text{Compute\_Latency} ,\text{Mem\_Latency})
\end{gathered}
\end{equation}


\subsection{Energy Modeling}

Energy is modeled via data collected for a 40nm process using Accelergy~\cite{accelergy} and its Aladdin~\cite{aladdin} and CACTI~\cite{cacti} plug-ins. In our model, compute, register access, and DRAM access energy are constant per word, whereas SRAM access energy per word scales with the number of SRAM rows and columns. The specific energy per access (EPA) values for each component are in Table~\ref{table:gemmini}.
\begin{equation}
\begin{gathered}
  \text{Energy}(i)  =  \text{Accesses}(i)\times\text{EPA}(i)\\
\text{Energy}  =  \text{MACs}\times\text{EPA}_{PE}+\sum_{i\in M}{\text{Energy}(i)}  
\end{gathered}    
\end{equation}

\subsection{Composing Performance Metrics}
\label{sec:composing}
In this work, we target co-design of an accelerator for a given DNN model. Thus far, all calculations have been per-layer. To compute the minimal hardware requirement for a set of layers, we take the max over layers of each hardware parameter. To compute performance, for example via energy-delay product (EDP), we sum the energies and latencies of each layer, and multiply these sums at the end. For layers that appear multiple times, one mapping is generated and its energy and latency are each multiplied by the number of times the corresponding layer appears. By setting the total EDP value as the gradient descent loss term, we minimize EDP for the full model, rather than find mapping/hardware configuration that minimizes EDP for individual layers. Say a model consists of layers $l$.
\begin{equation}\label{eqn:loss-function}
\text{EDP}(model) = \left(\sum_{l \in model}{\text{Energy}}\right) \times \left(\sum_{l \in model}{\text{Latency}}\right)
\end{equation}

Due to the scalability of gradient descent, we are able to optimize this objective with respect to all mappings in parallel, rather than one mapping at a time. This forms a different optimization problem over mappings compared to two-loop searchers that optimize for the EDP of individual layers. The flexibility of the GD loss function also enables the user to weight layers differently, which could be explored in future work.

\subsection{Correlation to Timeloop}
To demonstrate that our model does not compromise accuracy in order to provide differentiability and interpretability, we compare the predictions generated by the \sys{} differentiable model to Timeloop~\cite{parashar2019timeloop} and Accelergy. Specifically, we evaluate 73 matrix multiplication and convolutional layers, each of which are mapped onto 100 randomly generated Gemmini configurations and sampled approximately evenly for a total of 10,000 mappings. Figure~\ref{fig:correlation} shows that the EDP results from our differentiable model are on average within 0.18\% of Timeloop, with 98.3\% of results within 1\% of Timeloop. For very small layers with very low energy usage, there is up to 12.0\% error.
For these very small layers, Timeloop uses a ceiling function to compute energy based on the number of blocks accessed in DRAM, whereas the \sys{} differentiable model computes energy from the number of elements accessed.
Apart from these small layers, high correlation is observed because \sys{} captures the same relationships between mapping and latency and energy performance as Timeloop does. However, Timeloop models these relationships as an iterative program while \sys{} manages to express them in a mathematical framework to enable the use of white-box optimization algorithms for DSE. 

\subsection{Real Hardware Performance Modeling}
\label{sec:gemmini-model}
\label{sec:gemmini-training} 

In general, analytical models do not completely capture hardware performance~\cite{gem5correlation, mcpatcorrelation}. 
Variations caused by specific implementation details and complex hardware-software interactions may be unknown to the designer or difficult to capture mathematically. One potential solution is to augment analytical models with learned surrogate models. Since many learned models, such as deep neural networks or polynomial regression models, are differentiable, \sys{} is particularly well-suited to work with such models.

In this case, we train a deep neural network to predict the difference between our analytical model's latency predictions for a layer and the real latency of Gemmini-RTL, evaluated using FireSim~\cite{firesim}. The model's inputs include the layer's dimensions, a mapping (represented as in Section~\ref{sec:mapping}), and a hardware configuration.
The model's architecture is similar to that of the model used in Mind Mappings~\cite{hegde2021mind}. It contains 7 hidden fully-connected layers and a total of 5737 parameters.  

\section{\sys{} Optimization}\label{sec:optimization}
Constructing a differentiable performance model allows \sys{} to optimize hundreds of parameters (tens per layer, times tens of layers) at once using gradient descent (GD). As seen by its use for training neural network models with up to billions of parameters, GD is a highly performant and scalable optimization method. 

\begin{table}[t]
\centering
\begin{tabular}{|c|c|}
\hline
\textbf{Temporal Tiling Factors} & GD \\ \hline
\textbf{Spatial Tiling Factors} & GD \\  \hline
\textbf{Spatial Tiling Dimensions} & Constant \\ \hline
\textbf{Tensor Bypass} & Constant \\ \hline
\textbf{Loop Ordering} & Exhaustive \\
\hline
\end{tabular}
  \caption{Search algorithms for different design decisions.}
  \label{tab:search_algo}
  \vspace{-12pt}
\end{table}

\subsection{Search Strategy}
Table~\ref{tab:search_algo}
summarizes the search algorithms used by \sys{} to explore different mapping and design decisions. To determine the temporal and spatial tiling factors for each layer in the network (with 20 variables per layer), \sys{} employs GD. The GD loss term is the total performance metric, whose construction is described in Section~\ref{sec:composing}. Differentiability is implemented using PyTorch automatic differentiation. GD start points are generated via random hardware configuration, plus CoSA~\cite{cosa2021huang} mappings. 

We fix the spatial tiling dimensions (dataflow) decision to a weight stationary C(input channel)-K(output channel) mapping, as this is the primary spatial dataflow supported by the Gemmini generator.
Note that it is possible for the dataflow decision to be incorporated into the differentiable performance model, similar to the spatial tiling factor decision, by allowing the spatial factors from all problem dimensions to exceed 1.

For the current bypassing setup, we allocate one level of buffers for tensors with different data precisions, specifically the scratchpad for inputs and weights and accumulator for outputs.

\subsection{Loop Ordering}
\label{sec:loop-ordering}
We present two potential strategies for searching loop orderings in this section, which are compared in Section \ref{sec:loop-ordering-eval}.

\subsubsection{Iterative Loop Ordering Optimization}
For the iterative optimization strategy, depicted in Figure~\ref{fig:loop-ordering-iter}, we shuffle the order of loops for each layer every time mappings are rounded to the nearest valid mapping as explained in Section \ref{sec:mapping}. This typically occurs after several hundred gradient descent steps.
We select between three loop orderings per layer per level, each minimizing the data accesses for weights, outputs, and inputs respectively. We call these weight-stationary (WS), input-stationary (IS), and output-stationary (OS) orderings.
The differentiable model-predicted loop ordering that minimizes overall EDP, as in Equation~\ref{eqn:loss-function}, is selected.

\subsubsection{Gradient-Based Loop Ordering Optimization with Softmax Weighting}
\begin{figure}[t]
\centering
  \begin{subfigure}[t]{0.46\linewidth}
  \includegraphics[width=\linewidth, trim={0cm 1cm 13cm 0cm}]{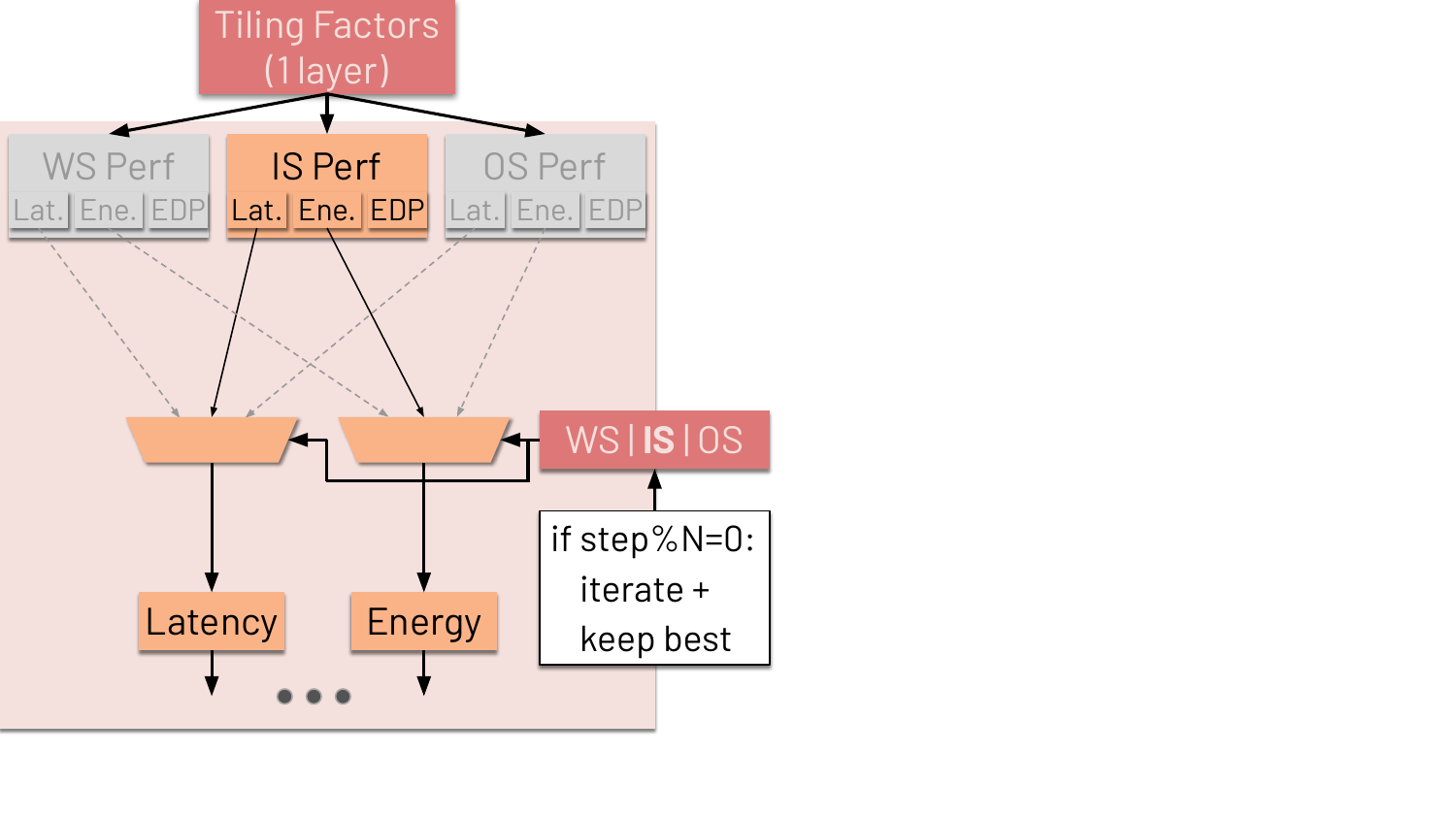}
  \caption{Iterative optimization, with IS loop ordering currently selected.}\label{fig:loop-ordering-iter}
  \end{subfigure}
  \hfill
  \begin{subfigure}[t]{0.46\linewidth}
  \includegraphics[width=\linewidth, trim={0cm 1cm 13cm 0cm}]{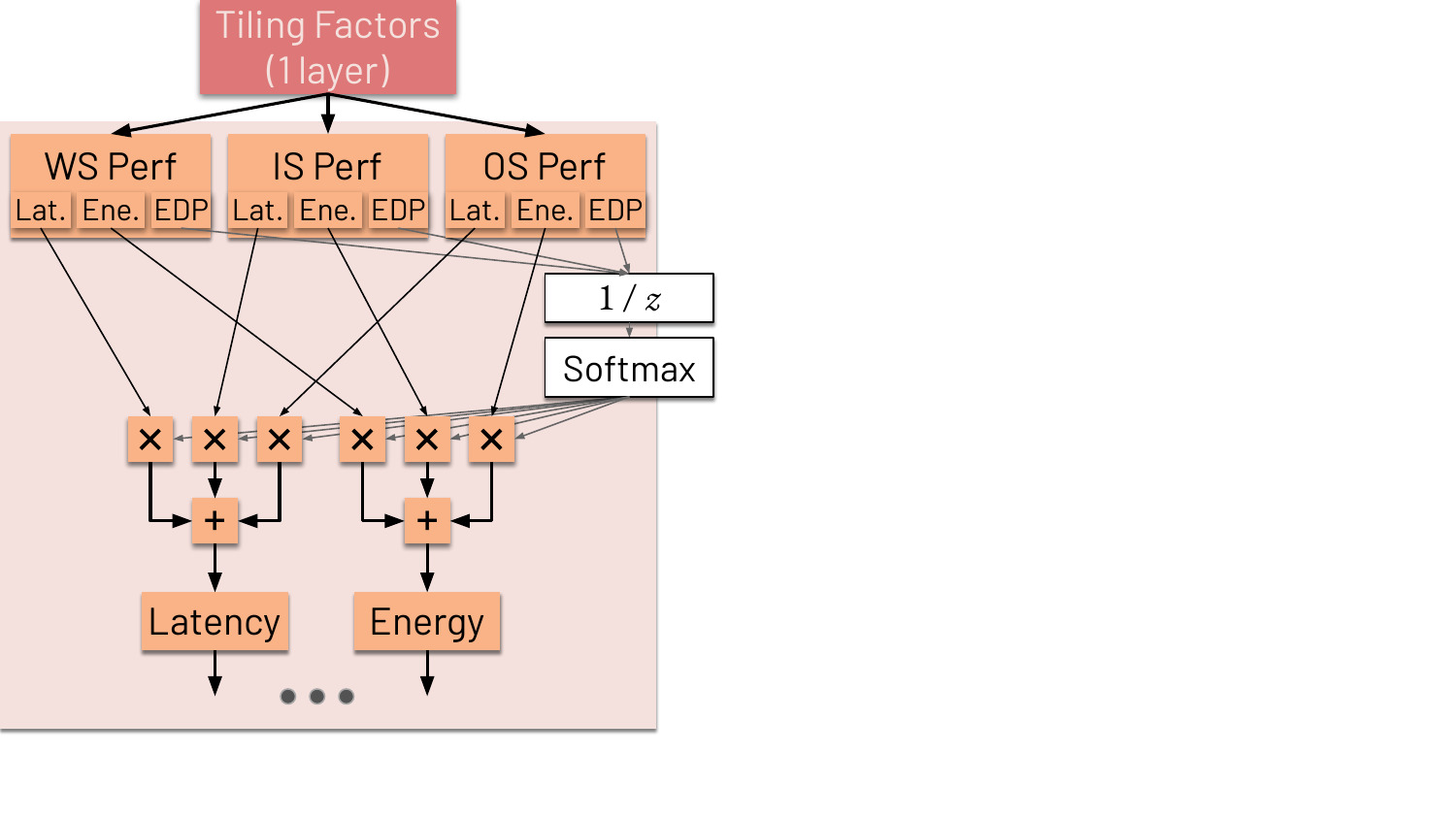}
  \caption{Gradient-based optimization with Softmax weighting.}\label{fig:loop-ordering-fig}
  \end{subfigure}
  \caption{\shepherding{Energy and latency prediction flow different loop ordering optimization schemes.}}
\end{figure}

The second optimization strategy involves integrating loop ordering into the gradient descent-based search by modifying the loss function. Like with iterative optimization, we consider WS, IS, and OS loop orderings for each level, for each layer. At every gradient descent step, we now consider the latency and energy of all loop ordering options and take them into account when updating tiling factors. We do so by combining the latency and energy predictions for each ordering using the softmax function $\sigma(z_i) = \frac{e^{z_{i}}}{\sum_{j=1}^K e^{z_{j}}}$. The case where loop ordering is flexible for one level is depicted in Figure~\ref{fig:loop-ordering-fig}. We first construct a vector of energies for each loop ordering options and a similar vector of latencies.
\begin{equation}
\begin{gathered}
\vec{\text{E}_l} = \begin{bmatrix}\text{Energy}_{l, WS} & \text{Energy}_{l, IS} & \text{Energy}_{l, OS}\end{bmatrix} \\
\vec{\text{L}_l} = \begin{bmatrix}\text{Latency}_{l, WS} & \text{Latency}_{l, IS} & \text{Latency}_{l, OS}\end{bmatrix}
\end{gathered}
\end{equation}

We then compute a vector $\vec{w_l}$ for each layer by applying the softmax function over the inverse EDPs of each loop ordering option. EDP is inverted so lower EDPs result in greater values in $\vec{w_l}$.
\begin{equation}
\vec{w_l} = \sigma\left(\frac{1}{\vec{\text{E}_l} \odot \vec{\text{L}_l}}\right)
\end{equation}



We use $\vec{w_l}$ to weight the energy and latency values of each loop ordering option before combining the energies and latencies of all layers in the model. The new loss function is as follows:
\begin{equation}
\text{Loss} = \left(\sum_{l \in model}{\vec{w_l} \cdot \vec{\text{E}_l}}\right) \times \left(\sum_{l \in model}{\vec{w_l} \cdot \vec{\text{L}_l}}\right)
\end{equation}

With this new loss function, gradient descent passes a gradient through all paths where $\vec{w_l}$ is not equal to 0, meaning tiling factors are optimized with awareness of which loop ordering is optimal for the current tiling factors. $\vec{w_l}$ also prevents these additional gradients from hindering optimization by weighting the gradients of more performant loop orderings more heavily than gradients of less performant loop orderings. 

\subsection{Other Optimization Details}
\subsubsection{Start Point Rejection}
In subsequent iterations of start point generation, if a start point's differentiable model-predicted performance is more than 10$\times$ that of the best start point seen thus far, it is rejected and a new hardware configuration is selected.

\subsubsection{Rounding}
Since gradient descent may result in non-integer tiling factors, before any mapping is evaluated, it is rounded to the nearest valid mapping. This is done by rounding each tiling factor to the nearest divisor of its corresponding problem dimension, subject to the constraint that the rounding process does not cause the product of tiling factors for that dimension to exceed the total problem size. This process iterates from the innermost to the outermost memory level.

\subsubsection{Preventing Exploration of Invalid Mappings}
We do not include tiling factors at the outermost (DRAM) level as optimization targets, and instead infer them by dividing the total problem size at each dimension by the product of the rest of the tiling factors for that dimension. 
In order to prevent exploration of invalid tiling factors that are less than 1, a loss term is added:

\begin{equation}
\sum_{(k,i,d) \in \{S, T\} \times M \times D}{max(1-f_{k,i,d}, 0)}
\end{equation}


\subsubsection{Pipeline Fusion}
Multilayer pipeline fusion is a critical technique to enhance the performance of DNN models whose computation can be decomposed into parallel streams of sequential layers. It allows concurrent processing across DNN layers, leading to higher hardware utilization and increased throughput. 
\sys{} is able to optimize critical DSE decisions with multilayer pipelined mappings.
Specifically, \sys{} remains effective for determining numerical variables such as spatial/temporal tiling factors,
and finding the best compute/buffer sizes (using a mapping-first approach). However, \sys{} faces challenges in making discrete pipeline fusion decisions, particularly when deciding the number of layers to fuse, balancing larger intermediate buffers against recomputation, and allocating DRAM bandwidth to different layers.
In this work, we do not search the space of pipeline fused layers.
\begin{table}[t]
\centering
\begin{tabular}{|c|c|}
\hline
\textbf{Training Workloads} & \textbf{Target Workloads} \\
\hline
AlexNet~\cite{alexnet} & BERT~\cite{devlin2018bert} \\
ResNeXt-50-32x4d~\cite{Xie2016resnext} & ResNet-50~\cite{resnet} \\
VGG-16~\cite{vgg16} & RetinaNet~\cite{retinanet} \\
DeepBench~\cite{deepbench} & U-Net~\cite{unet} \\
(OCR and Face Recognition) & \\
\hline
\end{tabular}
\caption{Training workloads for DNN-based performance prediction (Sections~\ref{sec:gemmini-training} and \ref{sec:gemmini_eval}) and target workloads on which we evaluate \sys{} (Section~\ref{sec:evaluation}). RetinaNet performance is evaluated on layers that are not part of its ResNet backbone. \label{table:test_workloads}\label{table:train_workloads} } 
\vspace{-12pt}
\end{table} 

 \begin{figure}[t]
 \centering
  \begin{subfigure}[b]{0.49\linewidth}
    \includegraphics[width=\linewidth]{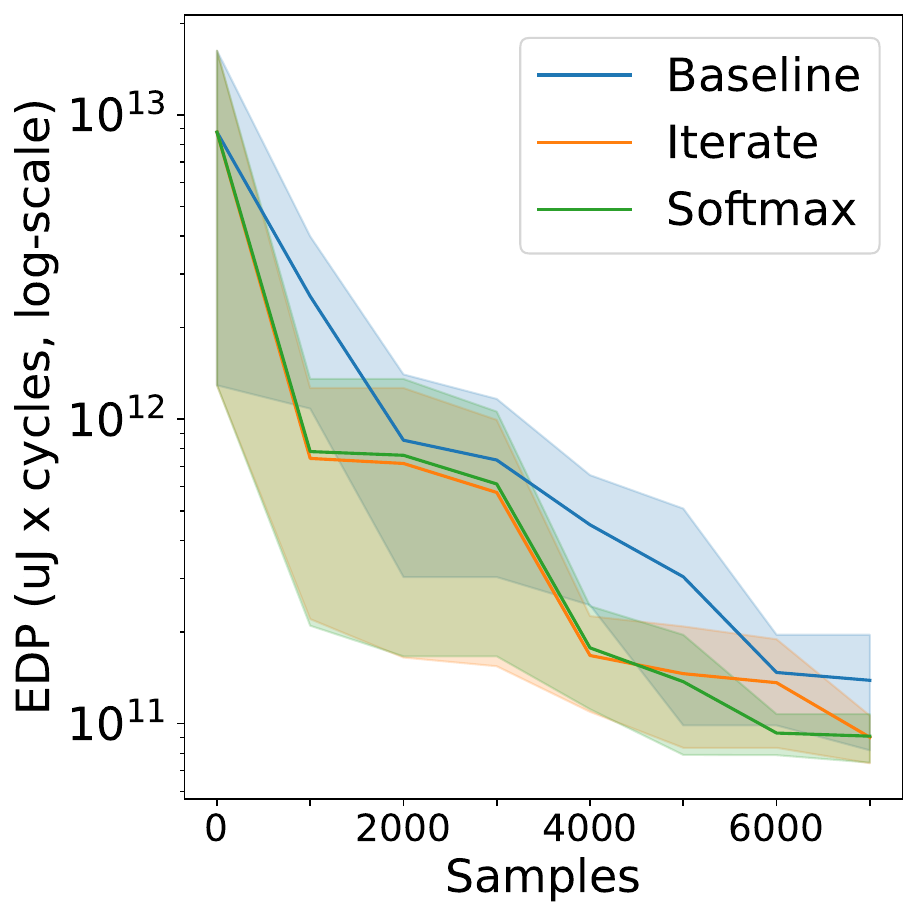}
        \caption{ResNet-50}
  \end{subfigure}
  \begin{subfigure}[b]{0.49\linewidth}
    \includegraphics[width=\linewidth]{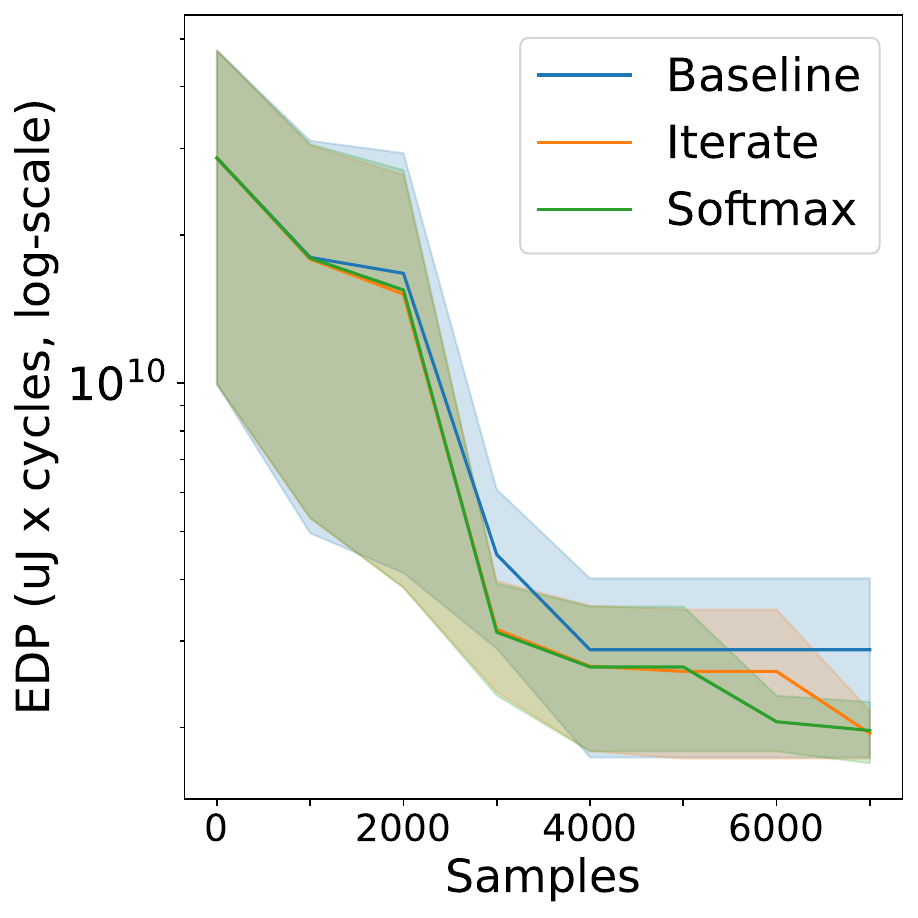}
        \caption{BERT}
  \end{subfigure}
  \caption{\shepherding{Comparison of no loop ordering optimization by \sys{} ("Baseline"), iterating over loop orderings every time mappings are rounded ("Iterate"), and gradient-based loop ordering ("Softmax"). The shaded regions represent a 95\% confidence interval across 3 runs. }}
  \label{fig:loop-ordering-eval}
\end{figure}

 \begin{figure*}[t]
  \centering
  \begin{subfigure}[b]{0.245\linewidth}
    \includegraphics[width=\textwidth]{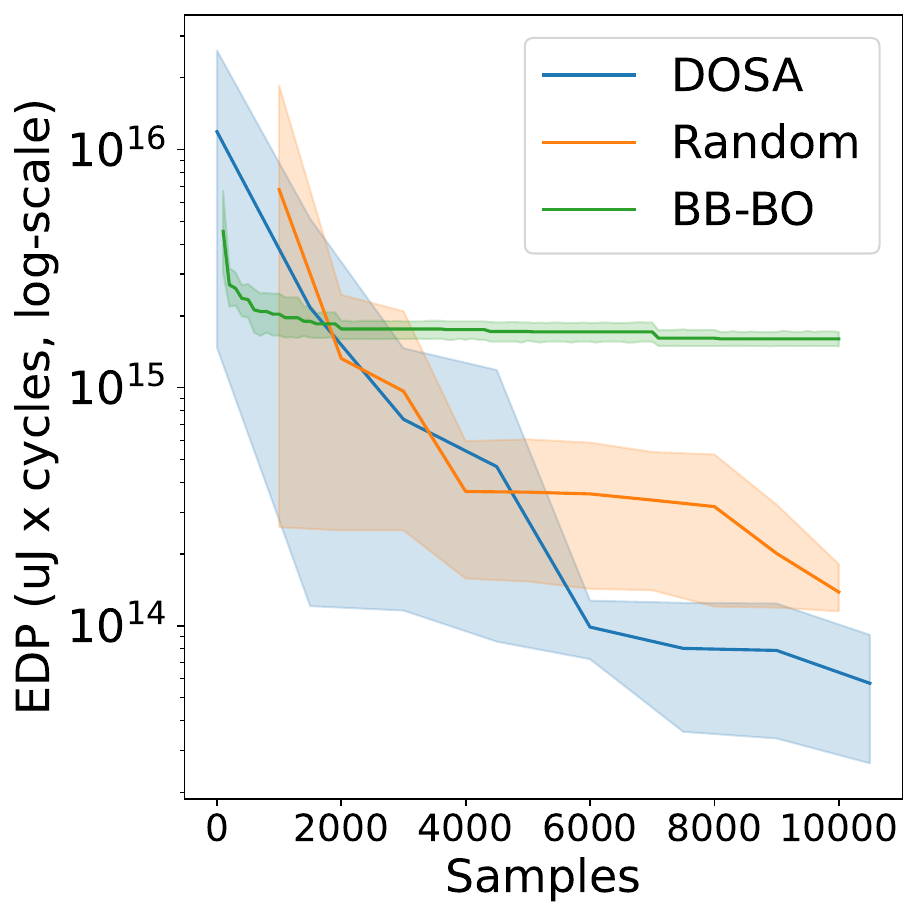}
        \caption{U-Net}
  \end{subfigure}
  \begin{subfigure}[b]{0.245\linewidth}
    \includegraphics[width=\textwidth]{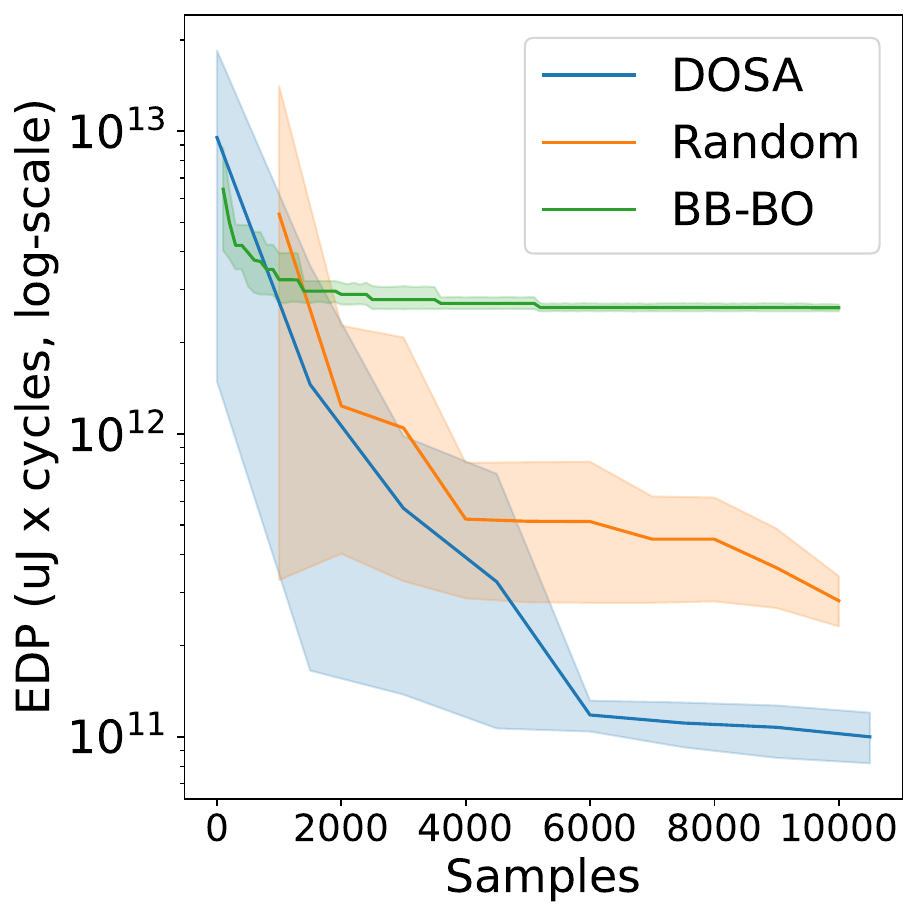}
        \caption{ResNet-50}
  \end{subfigure}
   \begin{subfigure}[b]{0.245\linewidth}
    \includegraphics[width=\textwidth]{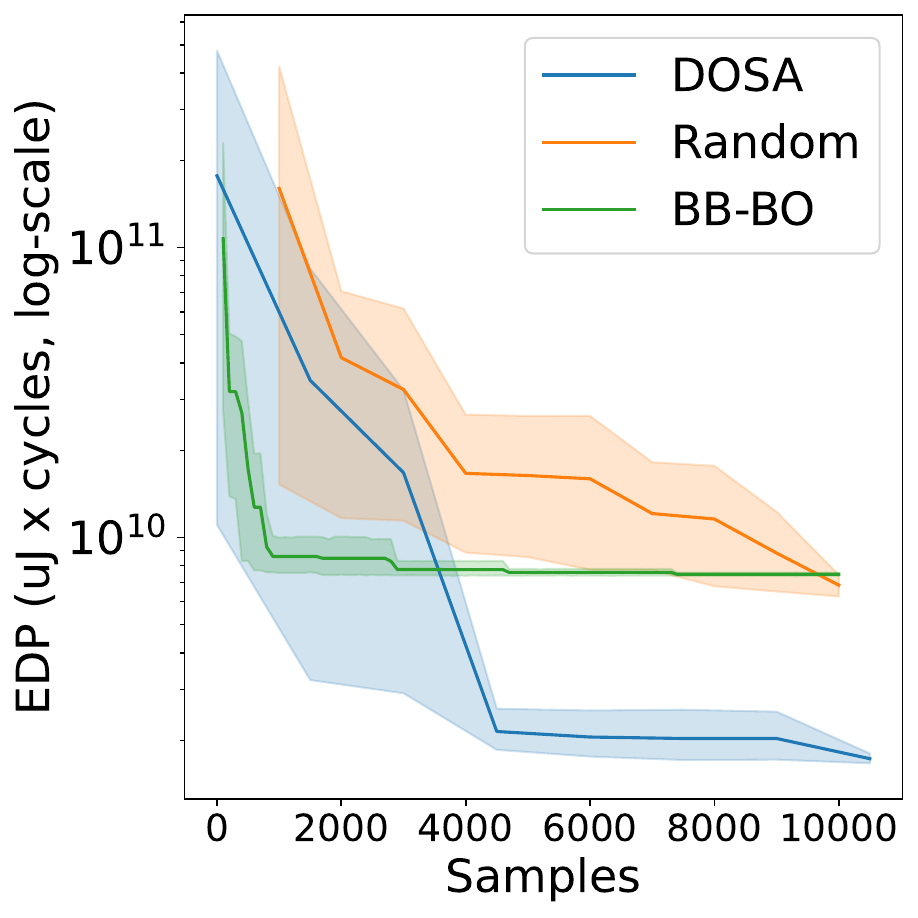}
        \caption{BERT}
  \end{subfigure}
  \begin{subfigure}[b]{0.245\linewidth}
    \includegraphics[width=\textwidth]{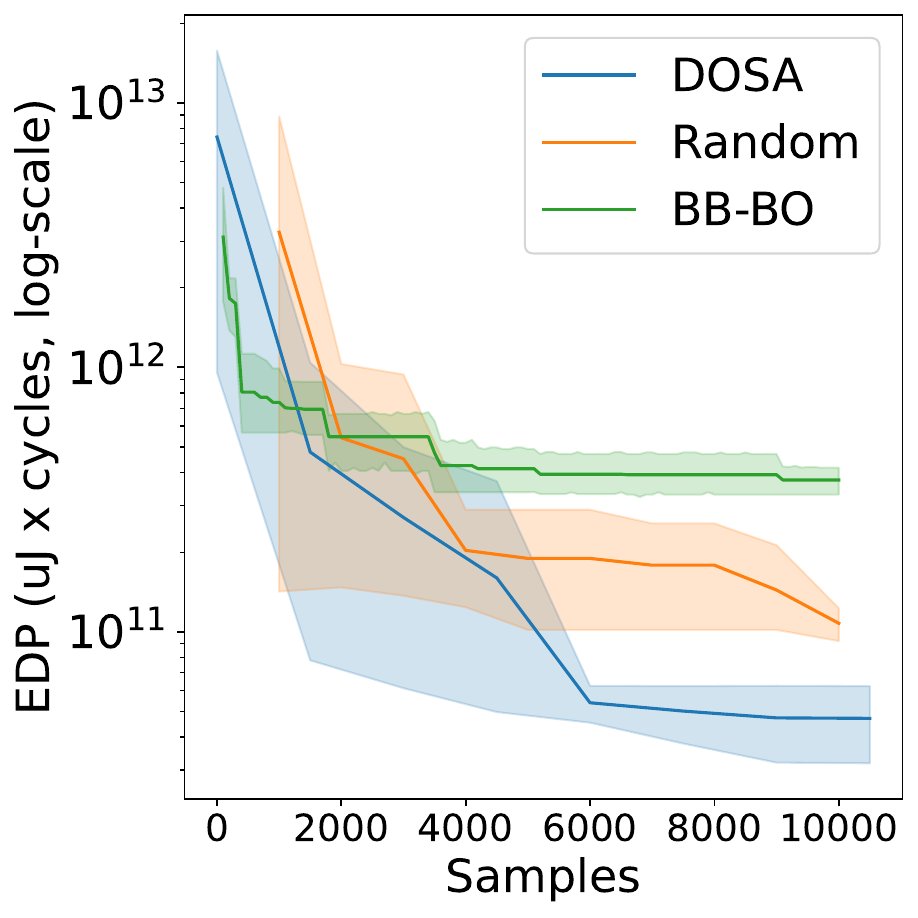}
        \caption{RetinaNet}
  \end{subfigure}

  \caption{\sys{} EDP optimization of Gemmini-TL on 4 distinct workloads, versus baselines. Each line represents the mean (across 5 runs) best point found after $x$ model evaluations. 
  The shaded regions represent a 95\% confidence interval across 5 runs. 
  }
  \label{fig:search_curve}
\end{figure*} 

 \begin{figure*}[t]
   \centering

  \begin{subfigure}[b]{0.245\linewidth}
    \includegraphics[width=\textwidth]{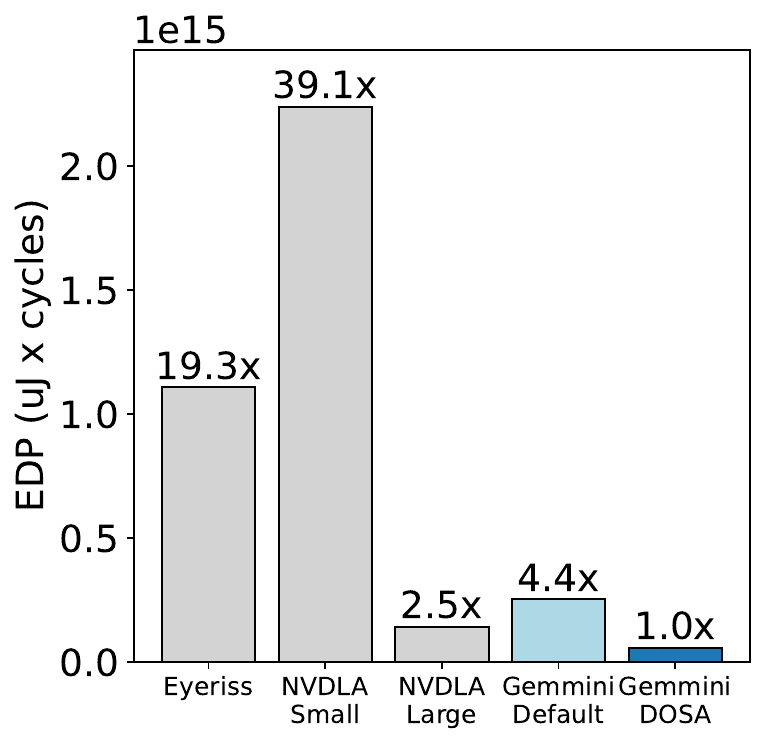}
        \caption{U-Net}
  \end{subfigure}
  \begin{subfigure}[b]{0.252\linewidth}
    \includegraphics[width=\textwidth]{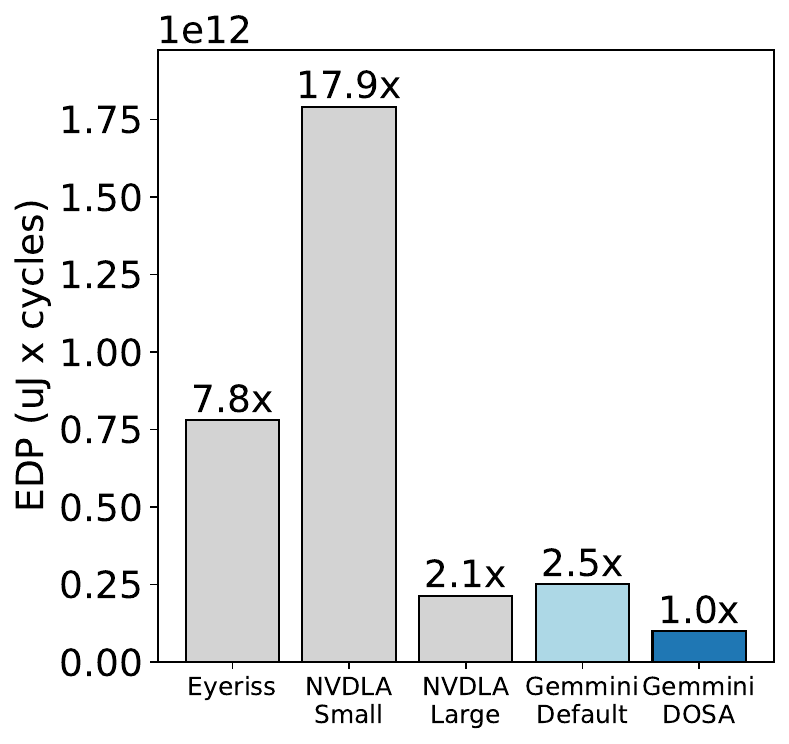}
        \caption{ResNet-50}
  \end{subfigure}
   \begin{subfigure}[b]{0.237\linewidth}
    \includegraphics[width=\textwidth]{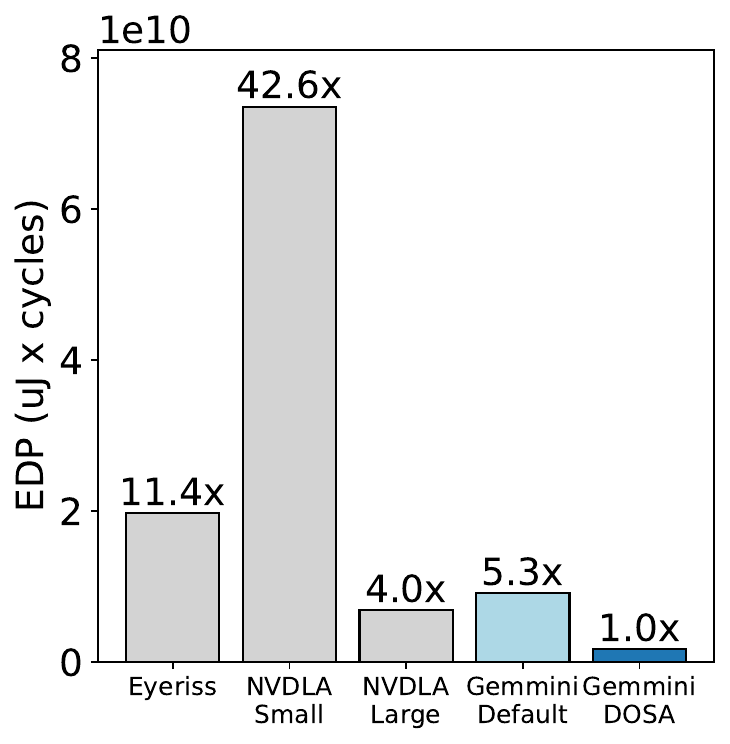}
        \caption{BERT}
  \end{subfigure}
  \begin{subfigure}[b]{0.245\linewidth}
    \includegraphics[width=\textwidth]{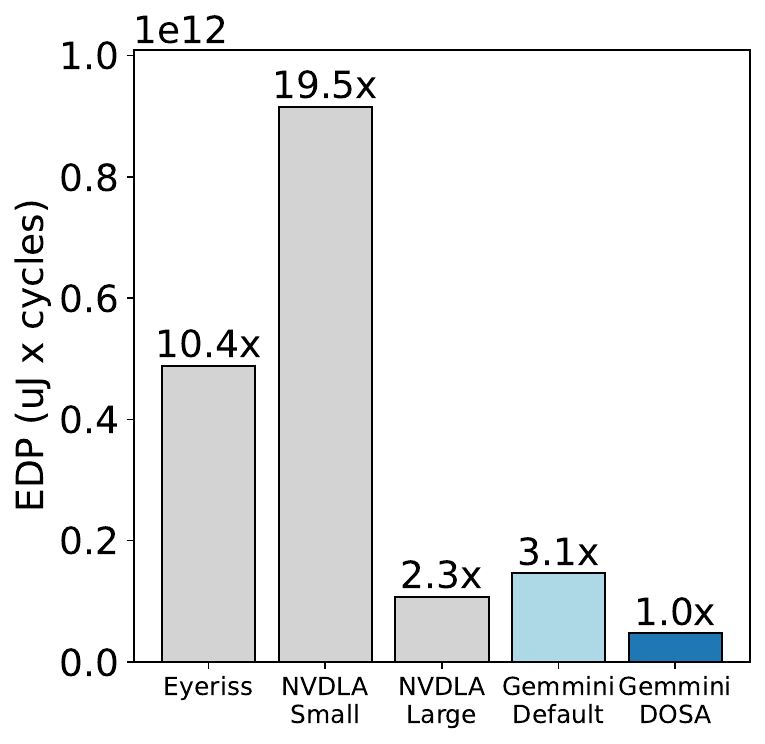}
        \caption{RetinaNet}
  \end{subfigure}

  \caption{\rebuttal{Energy-delay product (EDP) of baseline accelerators, compared to \sys{}-optimized Gemmini-TL. Bar labels represent EDP normalized to Gemmini-TL \sys{}.}
  }
  \label{fig:arch_compare}
\end{figure*} 


\section{Evaluation}

\label{sec:evaluation}

To differentiate between Timeloop performance evaluations and cycle-accurate evaluations of Gemmini RTL, from this point onward we use Gemmini-TL to refer to the Timeloop architectural definition of an accelerator analogous to Gemmini, and Gemmini-RTL to refer to the RTL implementation of Gemmini\footnote{Available at \url{https://github.com/ucb-bar/gemmini}.}. 
In this section, we first analyze the performance of an accelerator analogous to Gemmini using Timeloop simulation (Gemmini-TL), then demonstrate the ability of \sys{} to transfer to Gemmini-RTL.

\subsection{Experimental Setup}
We compare the performance of DSE algorithms on a variety of target DNN models that can handle a diverse set of tasks, such as natural language processing, image classification, object detection, and image segmentation. These target models are listed in Table~\ref{table:test_workloads}.
The hardware parameters we select using the capacity requirement calculations in Section~\ref{sec:methodology} are PE dimensions, accumulator SRAM sizing, and scratchpad SRAM sizing. PE dimensions come from spatial tiling factors, which can be directly used as they are always positive integers. PE array size is capped at 128x128. SRAM sizes are rounded up to increments of 1 KB.
For these experiments, the specific descent algorithm \sys{} uses is Adam, an optimizer similar to gradient descent with momentum. In Section~\ref{sec:loop-ordering-eval}, we use 7 start points, rounding happens every 300 steps, and GD is run for 890 steps on each start point. In Sections~\ref{sec:optimization-perf}--\ref{sec:gemmini_eval}, we use 7 start points, rounding happens every 500 steps and GD is run for 1490 steps on each start point. 

In addition to using CoSA to initialize GD start points, we apply it as a constant mapper to separate the effects of hardware and mapping improvements. CoSA requires a fixed partitioning for buffers that contain multiple tensors---we set up CoSA to partition the scratchpad equally between inputs and weights. Our Bayesian optimization-based hardware-mapping optimizer is a two-loop method which trains a Gaussian process model with 100 hardware designs and 100 mappings per layer per hardware design, and uses this model to select the hardware design and mappings with the best predicted performance from 1000 candidates per problem. We select these hyperparameters based on  Spotlight~\cite{sakhuja2023spotlight}. Finally, the random search baseline evaluates 10 hardware designs with 1000 mappings per layer per hardware designs.

In Sections~\ref{sec:loop-ordering-eval}--\ref{sec:separating}, we use Timeloop and Accelergy (with Aladdin and CACTI as plug-ins) to evaluate latency and energy. In Section~\ref{sec:gemmini_eval} we use RTL simulation in FireSim to evaluate latency, and Timeloop and Accelergy to evaluate energy.

\subsection{Evaluating Loop Ordering Optimization Strategies}
\label{sec:loop-ordering-eval}
We evaluate iterative loop ordering optimization and gradient-based loop ordering optimization on a subset of target workloads, specifically ResNet-50 and BERT. Each method uses the same start points. As shown in Figure~\ref{fig:loop-ordering-eval}, we find that gradient-based loop ordering optimization finds 1.58$\times$ better design points after 7000 samples compared to not searching loop orderings at all, whereas iterative optimization improves EDP by 1.70$\times$ over the baseline. Potential performance gains from searching loop orderings seem to be realized at similar levels by both methods, but slightly better by iteration after rounding. Iterative loop ordering optimization also requires significantly less computation. We use iterative loop ordering optimization for the experiments that follow.

\subsection{Hardware-Mapping Co-Search Performance}
\label{sec:optimization-perf}
Our evaluation finds that \sys{} is able to identify significantly more performant co-design points than either random search or Bayesian optimization with a similar number of samples. BB-BO uses Timeloop simulation as a black-box optimization metric for Gemmini-TL. The random search- and \sys{}-generated co-design points are also evaluated under this setup. After around 10,000 model evaluations, the geometric mean of EDP improvements for \sys{} versus random search is 2.80$\times$, and 12.59$\times$ versus BO. Evaluations done using Timeloop are considered equivalent to evaluations done using \sys{}'s differentiable model.

We find that in the regime of roughly less than 1000 samples, BB-BO performs best, likely because it performs more hardware search. However, only using 100 samples per hardware design may limit the full exploitation of each explore hardware design point, as by the 5,000 to 10,000 sample regime, BB-BO is overtaken by random search and \sys{}. BB-BO exhibits the least variance, while random search and \sys{} exhibit relatively high variance. However, \sys{} does tend to converge after around 6,000 samples, perhaps as it reaches close to optimal EDP.

\rebuttal{
In Figure \ref{fig:arch_compare}, we compare Gemmini-TL, with hardware and mappings optimized by DOSA, with other expert-optimized accelerator baselines (Eyeriss, NVDLA Small, NVDLA Large, and Gemmini default) for the four target workloads. We evaluate these accelerators using Timeloop and search 10,000 valid mappings per layer using the Timeloop random-pruned mapper. Gemmini-TL configurations searched by \sys{} consistently outperform the baselines by more than 2$\times$ in EDP.
}

\begin{figure}[t]
\centering
    \includegraphics[width=0.75\columnwidth]{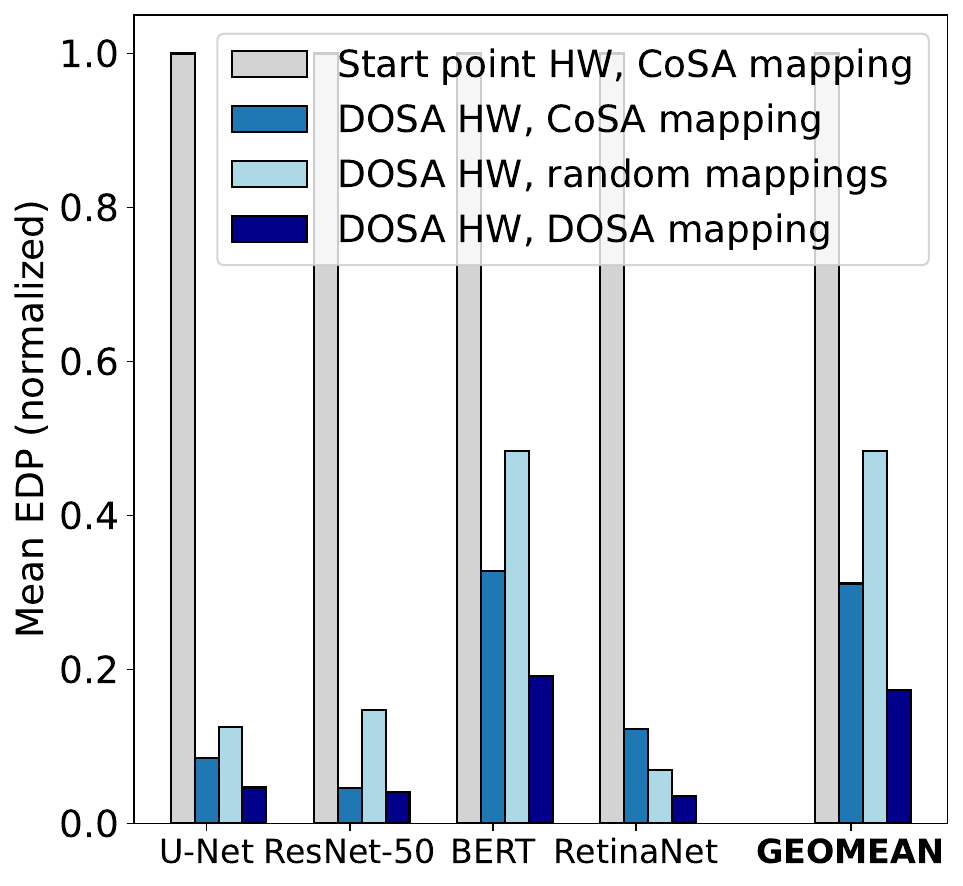}
    \caption{\sys{} improves performance under a constant mapper and produces near-optimal mappings for the hardware design points it selects.}\label{fig:start-points}
\end{figure}

\subsection{Separating the Effects of Hardware and Mapping Search}
\label{sec:separating}
We further find that \sys{} identifies both performant hardware design points and performant mappings. We run gradient descent 10 times, and use Timeloop to compare EDP at the GD start point (randomly selected hardware design with CoSA mappings) to EDP at the GD end point (\sys{}-generated hardware designs and mappings). We also evaluate these \sys{}-generated hardware designs with CoSA mappings, to see whether the hardware design improves under a constant mapper. This case study shows that \sys{} produces a 5.75$\times$ improvement over start point performance (geomean over 4 workloads, 10 GD instances each). Furthermore, with CoSA as a constant mapper, \sys{} generated hardware designs show a 3.21$\times$ improvement. This shows that end point hardware designs are better than start point hardware designs and that the performance improvements gained by \sys{} are not simply due to the use of a performant mapper (CoSA) in the loop. 

\sys{} mappings also show a 1.79$\times$ improvement over CoSA and a 2.78x improvement over a 1000-sample random mapper on the same hardware, showing that \sys{} identifies near-optimal mappings, competitive with or beating a state-of-the-art mapper, on the hardware design points it generates. The results per workload are shown in Figure~\ref{fig:start-points}.

\begin{figure*}[t]
\begin{minipage}{\textwidth}
\begin{minipage}{\textwidth}
  \centering
   \begin{subfigure}[b]{0.32\linewidth}
    \includegraphics[width=\textwidth]{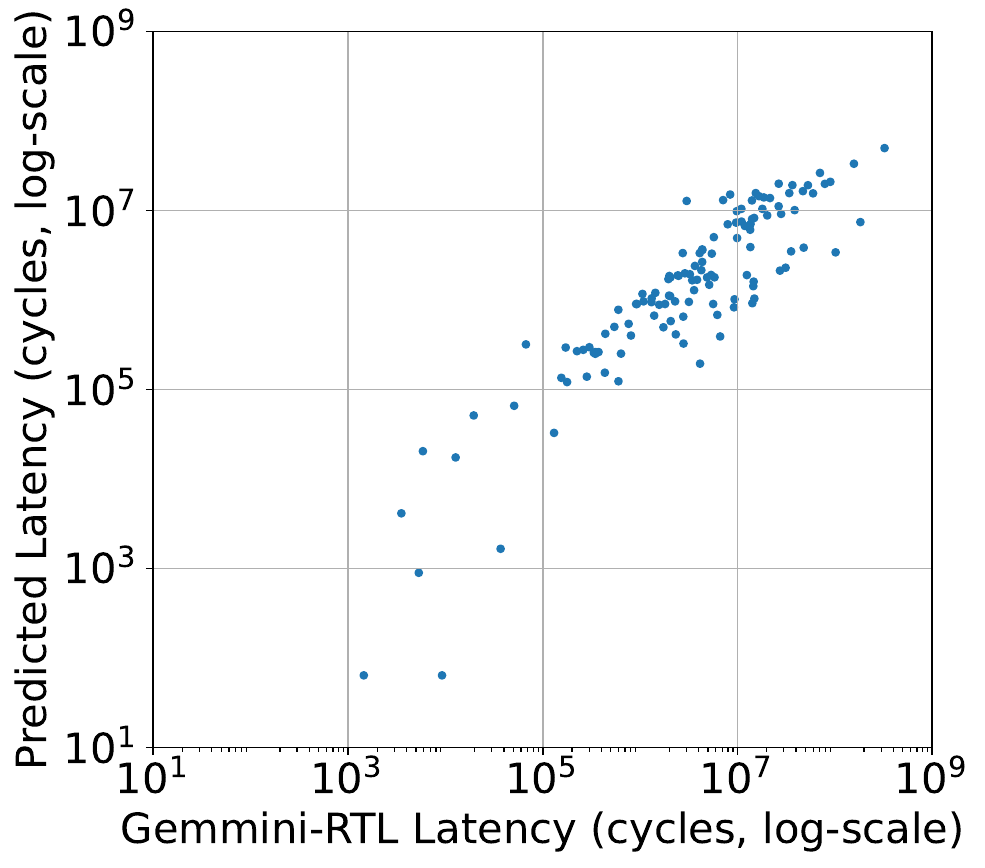}
        \caption{Analytical Only,\\corr.=0.87}\label{fig:gemmini-testset-analytical}
  \end{subfigure}
  \hspace{0.0\linewidth}
  \begin{subfigure}[b]{0.32\linewidth}
    \includegraphics[width=\textwidth]{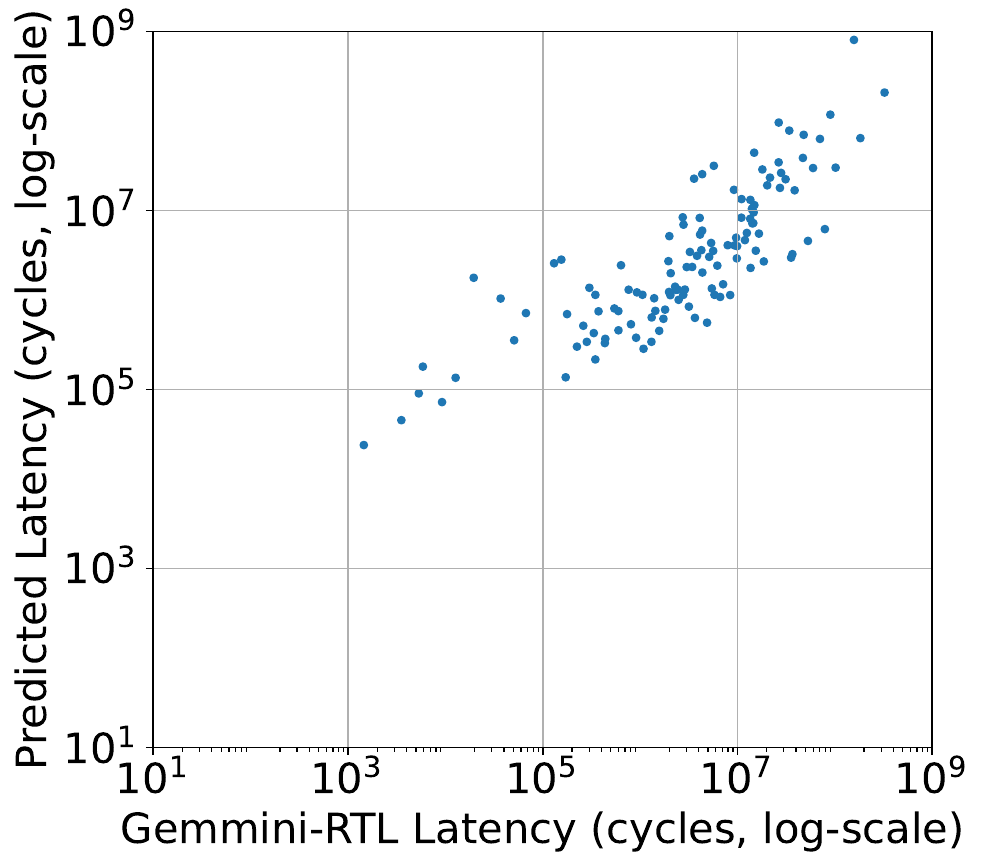}
        \caption{DNN Only,\\corr.=0.84}
  \end{subfigure}
  \hspace{0.0\linewidth}
  \begin{subfigure}[b]{0.32\linewidth}
    \captionsetup{justification=centering}
    \includegraphics[width=\textwidth]{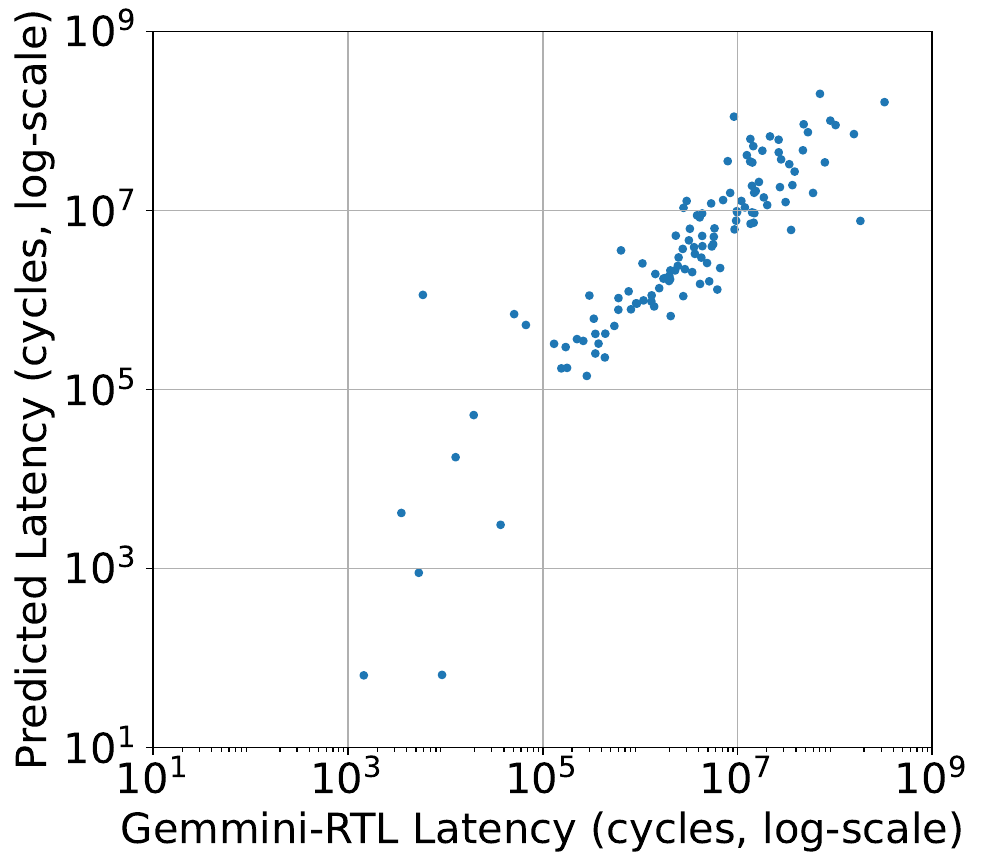}
        \caption{Analytical + DNN,\\corr.=0.92}
  \end{subfigure}
  \caption{\rebuttal{Accuracy of Gemmini-RTL latency models on test split of random mappings (training workloads from Table~\ref{table:train_workloads}, unseen mappings). Correlation metric is Spearman rank correlation.}}
  \label{fig:gemmini-dnn} \vspace{8pt}
\end{minipage} \newline \hspace{0.0\linewidth} 
\begin{minipage}{\textwidth}
  \centering
   \begin{subfigure}[b]{0.32\linewidth}
    \includegraphics[width=\textwidth]{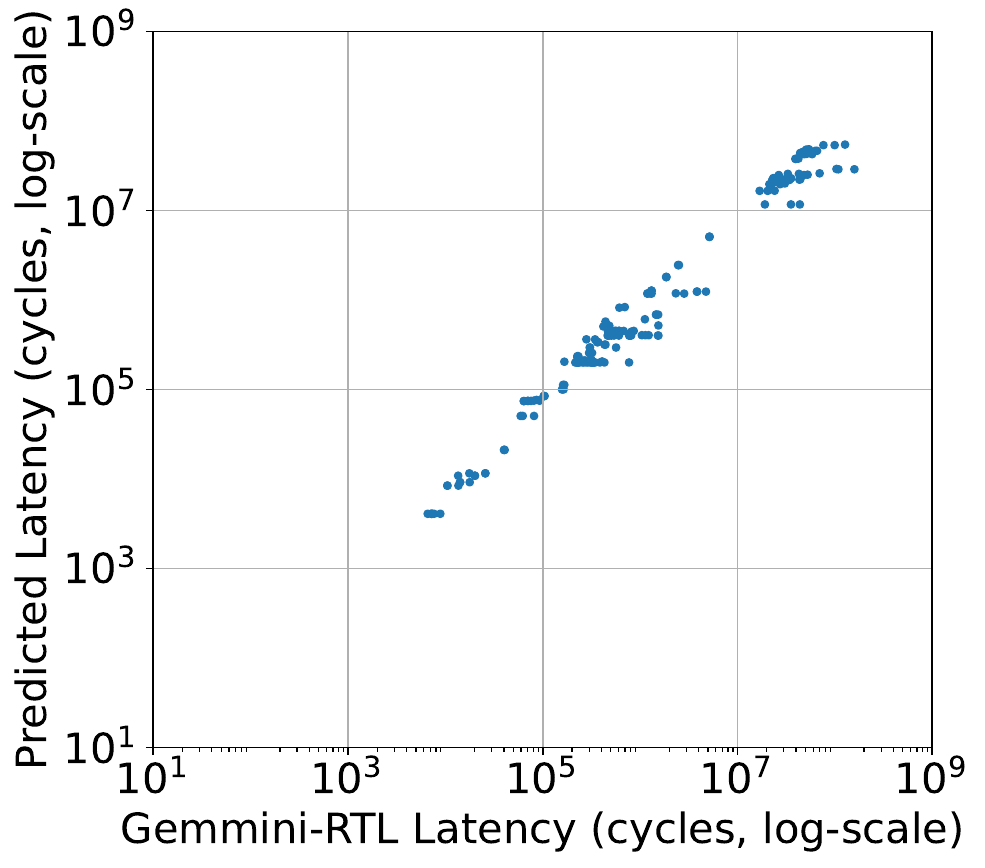}
        \caption{Analytical Only,\\corr.=0.97}
  \end{subfigure}
  \hspace{0.0\linewidth}
  \begin{subfigure}[b]{0.32\linewidth}
    \includegraphics[width=\textwidth]{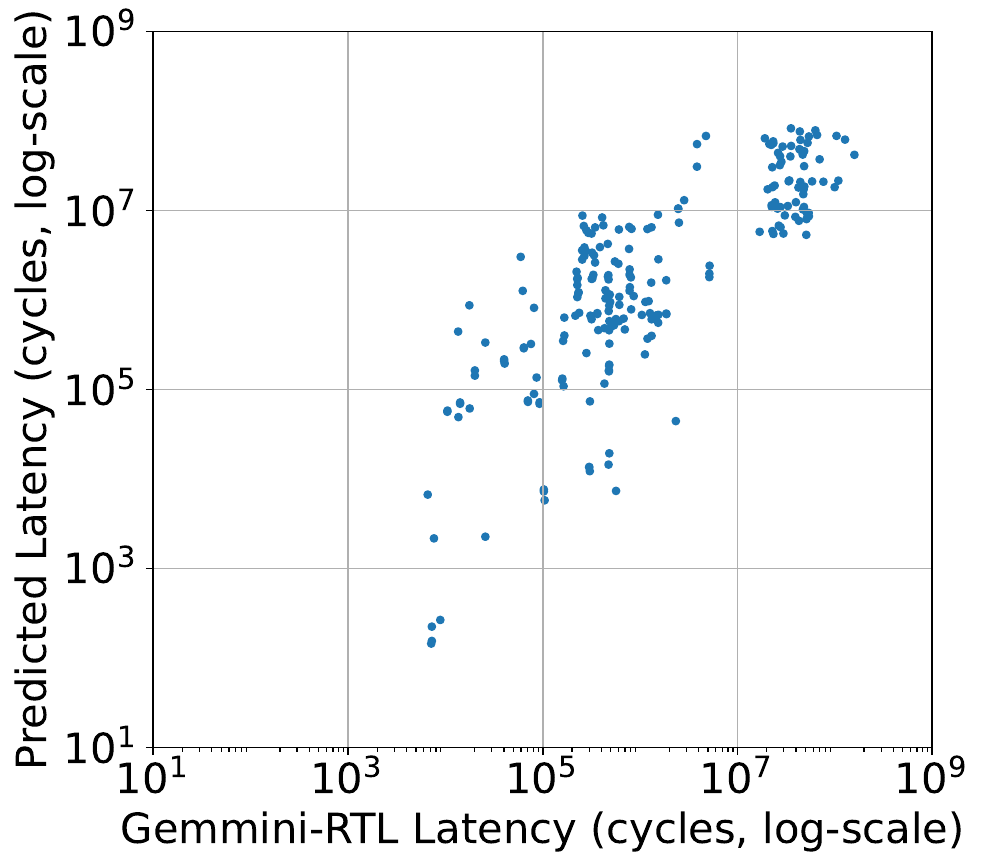}
        \caption{DNN Only,\\corr.=0.79}\label{fig:gemmini-dosa-dataset-dnn}
  \end{subfigure}
  \hspace{0.0\linewidth}
  \begin{subfigure}[b]{0.32\linewidth}
    \captionsetup{justification=centering}
    \includegraphics[width=\textwidth]{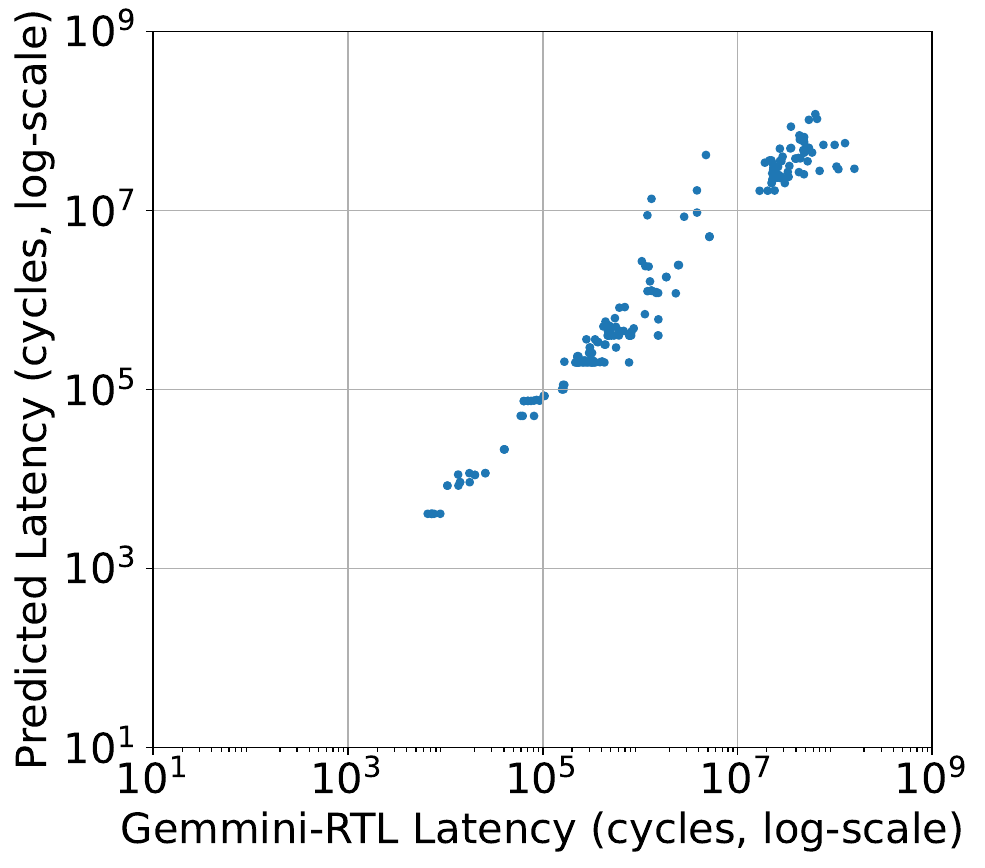}
        \caption{Analytical + DNN,\\corr.=0.97}
  \end{subfigure}
  \caption{\rebuttal{Accuracy of Gemmini-RTL latency modeling on mappings generated using \sys{}. These are mappings for the target workloads in Table~\ref{table:test_workloads}, which are not included in DNN training data.}}
  \label{fig:gemmini-dnn-not-training-set}
\end{minipage}
\end{minipage}
\end{figure*}

\begin{figure}
    \includegraphics[width=0.75\columnwidth]{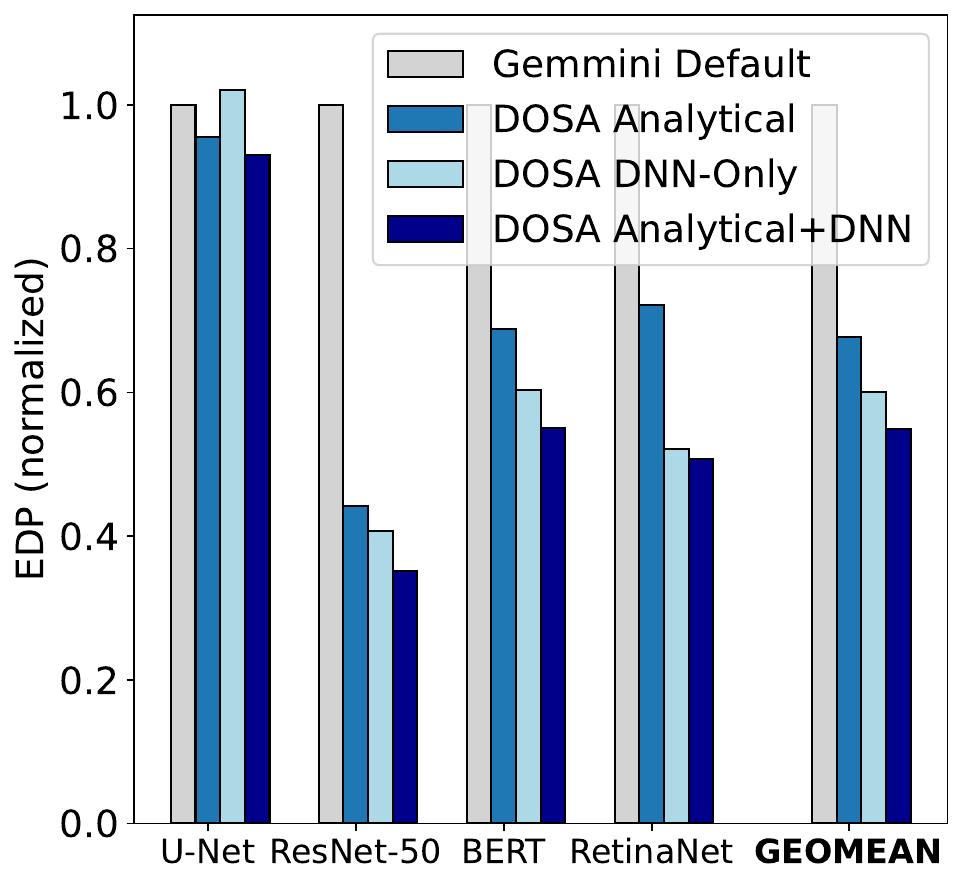}
    \caption{\rebuttal{\sys{} optimization of Gemmini-RTL using various performance models, compared to Gemmini's hand-tuned hardware and mapper.}}\label{fig:gemmini-results}
\end{figure}

\subsection{Gemmini-RTL Optimization with \sys{}}\label{sec:gemmini_eval}

In this section, we assess the efficacy of our one-loop differentiable-model-based gradient descent approach for real hardware design. 
We explore three potential approaches for latency modeling of Gemmini-RTL: an \emph{analytical-only} approach using the model described in Sections~\ref{sec:resource_req}--\ref{sec:composing},
a \emph{DNN-only} model trained from scratch from measured Gemmini-RTL performance data and our \emph{DNN-augmented analytical model} approach from Section~\ref{sec:gemmini-training}. 
Note that energy is predicted using the \sys{} differentiable analytical model in all cases.

\subsubsection{DNN Model Training Setup}
We utilize FireSim to generate cycle-accurate Gemmini-RTL latencies on our training set of models (Table~\ref{table:train_workloads}). Specifically, we generate a relatively small dataset of 1567 random mappings, roughly evenly distributed among the layers in Table~\ref{table:train_workloads}. This dataset is used to train the two DNN performance models, which have the same architecture and are trained using the same hyperparameters. The models are trained for 50,000 epochs. The models could likely be trained to greater accuracy with a larger dataset, but we limit dataset size to more accurately represent a real hardware design environment. 

\subsubsection{Prediction Accuracy}
\label{sec:gemmini-pred-accuracy}

Figure~\ref{fig:gemmini-dnn} compares the accuracy of each approach when modeling Gemmini-RTL performance on unseen random mappings of workloads in the training set. To quantify prediction accuracy, we measure the Spearman rank correlation of each model's predictions. Spearman rank correlation measures the strength of the associations between two variables by assessing the monotonicity of the relationship between those variables~\cite{spearman}. The combined model providing the highest accuracy, with a correlation coefficient of 0.92. The analytical-only model is the next most accurate model on this dataset with a correlation coefficient of 0.87, and the DNN-only model is the least accurate by a small margin with a correlation coefficient of 0.84.


\subsubsection{Optimization Performance}
\label{sec:rtl-optimization}

To evaluate \sys{}'s real hardware optimization performance, we run \sys{} using the analytical-only, DNN-only and DNN-augmented latency models.
Specifically, for each latency model and for each target workload, we generate a predicted optimal set of mappings and buffer sizes for 16x16 PE Gemmini-RTL, fixing PE dimensions and adjusting only buffer sizing and mappings. \rebuttal{We fix PE dimensions to keep accelerator size in the same order of magnitude as default Gemmini-RTL. This allows for a more apples-to-apples comparison against the hand-tuned default design point, and ensures that generated configs can be simulated using FireSim.} We compare the performance of \sys{}-generated mappings to the mappings generated by the Gemmini-RTL default heuristic-based mapper, and the default scratchpad and accumulator sizings of 128 KB and 32 KB respectively (256 KB and 64 KB including double-buffering), which were selected using heuristics similar to those in Interstellar~\cite{interstellar-asplos2020}.
As mentioned above, Gemmini-RTL latency is evaluated using FireSim, while energy is evaluated using Timeloop and Accelergy, with CACTI as a plug-in. 

Figure~\ref{fig:gemmini-dnn-not-training-set} shows the prediction accuracy of each model on the mappings produced using \sys{} during these experiments. These are mappings for layers not present in the training dataset. On this dataset, the DNN-only model is clearly less accurate than the analytical-only model or the DNN-augmented analytical model, as seen by the outliers in Figure~\ref{fig:gemmini-dosa-dataset-dnn}. This reflects previous work \cite{kao2022demystifying}, which has shown that DNN-only methods for mapspace exploration \cite{hegde2021mind} have difficulty generalizing beyond their training sets. The analytical model, which is not fit to any particular workload, actually improves in prediction accuracy compared to the original test dataset (Figure~\ref{fig:gemmini-testset-analytical}), likely because this new dataset consists of performant mappings generated by \sys{}, and as such is more uniform.

Figure~\ref{fig:gemmini-results} shows the EDP of Gemmini-RTL after it is optimized  for each target workloads, using the three latency models. 
The analytical-only and DNN-only models respectively yield 1.48$\times$ and 1.66$\times$ improvements over Gemmini's default buffer sizes and tilings. Despite its drop-off in \emph{prediction accuracy}, the DNN-only model outperforms the analytical model in \emph{optimization performance} on the three workloads other than U-Net. U-Net contains weight sizes unseen in the training set (Table~\ref{table:train_workloads}), again demonstrating DNN models' difficulty in generalizing.

When DNN and analytical models are combined, we improve optimization performance even further, to 1.82$\times$ over default, while maintaining a level of prediction accuracy (on \sys{}-generated points) higher than that of the DNN-only model and similar to that of the analytical-only model. Unlike the DNN-only model, the DNN-augmented analytical model does not produce outliers in our experiments, since its outputs are constrained using the analytical model prediction. 

Table~\ref{table:optimal_gemmini_config} shows the buffer sizes selected by \sys{} with the DNN-augmented latency model. The buffer size ratios (scratchpad size divided by accumulator size) identified by \sys{} using the analytical+DNN model setup range from 1.28 to the original heuristically selected ratio of 4. We find that for all four target workloads, \sys{} sizes both the accumulator and scratchpad significantly larger than the default sizes, indicating that these buffers may be underprovisioned for such workloads. Furthermore, for the three convolutional neural networks (U-Net, ResNet-50, and RetinaNet), the ratio of scratchpad to accumulator size is smaller than in the default configuration.

This experiment demonstrates that \sys{}'s gradient descent-based optimization technique is compatible with performance models other than the analytical model presented in this work. In fact, \sys{} allows for a more iterative process when moving from simulation to real hardware, as the model for each objective (latency, energy, and in future work, potentially area) can be replaced and augmented independently as demonstrated here. Furthermore, the components that remain analytically modeled can be easily tuned as more accurate data becomes available or design decisions change. For example, energy-per-access numbers could be updated based on measured numbers once a process node is selected or modified, which would be orders of magnitude more efficient than generating large amounts of data for a DNN model to consume.


\begin{table}[t]
\centering
\small\tabcolsep3.5pt\begin{tabular}{|c|c|c|c|}
\hline
  \multicolumn{2}{|c|}{} & Accumulator (KB) & Scratchpad (KB) \\ \hline
  \multicolumn{2}{|c|}{\textbf{Gemmini Default}} & 32 & 128 \\ \hline
  \multirow{4}{*}{\begin{tabular}[c]{@{}c@{}} \textbf{DOSA-Optimized}\\\textbf{Gemmini-RTL} \end{tabular}} &
  U-Net & 123 & 322 \\ \cline{2-4}
  & ResNet-50 & 196 & 251 \\ \cline{2-4}
  & BERT & 64 & 256 \\ \cline{2-4}
  & RetinaNet & 112 & 261 \\
\hline
\end{tabular}
\caption{\rebuttal{Gemmini configurations generated by \sys{} Analytical+DNN.}}
\label{table:optimal_gemmini_config} \vspace{-12pt}
\end{table}

\section{Conclusion} 

In this paper, we present \sys{}, a differentiable model-based approach to mapping-first DSE. By constructing a differentiable analytical performance model for a DNN accelerator, we can use gradient descent to perform an efficient one-loop co-search of both the hardware and mapping spaces.
This enables us to to perform DSE targeting multi-layer neural net workloads, attaining an EDP 2.80$\times$ better than random search and 12.59$\times$ better than Bayesian optimization, while using a similar number of samples. Furthermore, we find that \sys{} not only improves hardware design performance by 3.21$\times$ under a constant mapper, but also beats a state-of-the-art mapper on the hardware designs it selects.

\sys{} demonstrates that interpretable, designer-trusted architectural modeling and ML-based optimization methods can be combined to improve the convergence of DSE. 
We pair our analytical latency model with a DNN model trained on RTL simulation data, improving EDP over the default Gemmini configuration 1.48$\times$ with just the analytical model and by 1.82$\times$ when analytical and learned models are combined.
With this work, we move one step closer to bridging the gap between architectural models and real silicon.

\begin{acks}
Many thanks to the anonymous MICRO reviewers and our shepherd for suggesting important additions.
We also thank Hasan Genc and Divija Hasteer for their help with Gemmini, and Xiangyu (Sam) Xu and
Jonathan Wang for their help generating baseline data.

This research is supported by NSF Award CCF-2238346 and by SLICE Lab industrial sponsors and affiliates Amazon, AMD, Apple, Google, Intel, and Qualcomm.
Any opinions, findings, and conclusions or recommendations expressed in this material are those of the author(s) and do not necessarily reflect the views of the National Science Foundation.
\end{acks}

\bibliographystyle{ACM-Reference-Format}
\bibliography{dosa-refs}

\clearpage
\appendix
\section{Artifact}
\label{app:artifact}

\subsection{Abstract}

This section describes how to access the artifacts for \sys{}, and run several of the experiments from Section~\ref{sec:evaluation}, in particular reproducing the results of Gemmini optimization using \sys{}. We reproduce experiments in both Timeloop and RTL simulation environments.

\subsection{Artifact check-list (meta-information)}
{\small
\begin{itemize}
  \item {\bf Run-time environment:} User machine, AWS FPGA Developer AMI 1.12.2.
  \item {\bf Hardware:} AWS EC2 instances (c5.4xlarge, f1.2xlarge).
  \item {\bf Metrics:} Accelerator energy-delay product. Evaluated on Timeloop architectural simulator and FireSim RTL simulation.
  \item {\bf Output:} Plots showing model accuracy, DSE sample efficiency, optimization performance. Energy and latency numbers for Gemmini hardware configurations and mappings.
  \item {\bf Experiments:} DSE sample efficiency and EDP of resulting designs compared to baselines. 
  \item {\bf How much disk space required:} 20 GB (on user machine), 200 GB (on EC2 instance).
  \item {\bf How much time is needed to prepare workflow:} 2 hrs.
  \item {\bf How much time is needed to complete experiments:} approximately 24 hrs, dependent on user machine.
  \item {\bf Publicly available:} Yes.
  \item {\bf Code licenses:} Multiple, see download.
  \item {\bf Archived:} \url{https://doi.org/10.5281/zenodo.8358253}
\end{itemize}
}

\subsection{Description}

\subsubsection{How to access}
The artifact includes several git repositories archived on Zenodo at \url{https://doi.org/10.5281/zenodo.8358253}.
\begin{itemize}
\item The code for \sys{}, stored in the archive \lstinline|dosa.zip|, plus a version with all submodules initialized under \lstinline|dosa-full.zip|.
\item The FireSim, Chipyard, and Gemmini configurations (hardware and software) used for \sys{}, archived under the names \lstinline|firesim-dosa.zip|, \lstinline|chipyard-dosa.zip|, and \lstinline|gemmini-dosa.zip| respectively.
\end{itemize}

\subsubsection{Hardware dependencies}
One AWS EC2 c5.4xlarge instance (“manager” instance), and one f1.2xlarge instance  ("run farm" instance) are required. The latter will be launched automatically by FireSim’s manager. We have provided a pre-built FPGA image to avoid the long latency (about 8 hours) of the FPGA build process.

\subsubsection{Software dependencies}
\sys{}, Timeloop, and Accelergy can be installed on most Linux-based user machines.

For FireSim-based experiments, use ssh or mosh on your local machine to remote access EC2 instances. All other requirements are automatically installed by scripts in the following sections.



\subsection{Installation}

\subsubsection{Installing \sys{}}
On a user machine with Python 3.10 or greater, clone the archived \sys{} code: 
\begin{lstlisting}
curl -Ls -w %{url_effective} -o a https://doi.org/10.5281/zenodo.8358253 > DL_url
wget $(cat DL_url)/files/dosa.zip
unzip dosa.zip
\end{lstlisting}

First, acquire a Gurobi optimizer license\footnote{\url{https://www.gurobi.com/features/academic-named-user-license/}} and download it to path of choice \lstinline|($license_path)|. Next, run the following:
\begin{lstlisting}
export GRB_LICENSE_FILE=($license_path)
cd dosa
pip3 install -e .
\end{lstlisting}

\subsubsection{Timeloop and Accelergy}
Install Timeloop and Accelergy on the user machine. The following dependencies are required (command provided for Debian-based systems):
\begin{lstlisting}
sudo apt install scons libconfig++-dev libboost-dev libboost-iostreams-dev libboost-serialization-dev libyaml-cpp-dev libncurses-dev libtinfo-dev libgpm-dev git build-essential python3-pip
\end{lstlisting}

Timeloop and Accelergy are available as submodules of the \sys{} repository. First, install Accelergy and its plug-ins. Within \lstinline|dosa|:
\begin{lstlisting}
git submodule update --init --recursive
cd accelergy-timeloop-infrastructure/src/accelergy
pip3 install .
cd ../cacti
make
cd ..
mv cacti ~/.local/bin/
cd ../accelergy-cacti-plug-in
pip3 install .
cd ../accelergy-aladdin-plug-in
pip3 install .
cd ../accelergy-table-based-plug-ins
pip3 install .
accelergy
accelergyTables
\end{lstlisting}

Install Timeloop and add its executables to your \verb|PATH|:
\begin{lstlisting}
cd ../timeloop/src
ln -s ../pat-public/src/pat .
cd ..
scons --accelergy --static -j4
export PATH=$PATH:$(pwd)/build
\end{lstlisting}

The Timeloop\footnote{\url{https://timeloop.csail.mit.edu/timeloop/installation}} and Accelergy\footnote{\url{https://timeloop.csail.mit.edu/accelergy/installation}} documentation may be helpful if any issues arise.

\subsubsection{FireSim-Based Experiments}
First, follow the instructions on the FireSim website~\footnote{\url{https://docs.fires.im/en/1.17.1/Getting-Started-Guides/AWS-EC2-F1-Getting-Started/index.html}} to create an EC2 manager instance. Complete the steps in the “AWS EC2 F1 Getting Started Guide”. Once you have completed up to and including "Setting up your Manager Instance / Key setup, Part 2" in the FireSim docs, you should have a manager instance set up, with an IP address and key. Use ssh or mosh to log in to the instance. Next, in \lstinline|/home/centos|, clone the archived FireSim repository. 
\begin{lstlisting}:
curl -Ls -w %{url_effective} -o a https://doi.org/10.5281/zenodo.8358253 > DL_url
wget $(cat DL_url)/files/firesim-dosa.zip
unzip firesim-dosa.zip
\end{lstlisting}

Run the following, which will initialize dependencies and set up FireSim and Chipyard:
\begin{lstlisting}
cd firesim-dosa
git checkout dosa
./build-setup.sh
sudo yum install autoconf
source sourceme-f1-manager.sh
firesim managerinit --platform f1
\end{lstlisting}

After sourcing, complete the steps in "Setting up your Manager Instance / Completing Setup Using the Manager". Note that \lstinline|sourceme-f1-manager.sh| must be sourced every time you log in to the instance.

Finally, get the FPGA image used for this experiment. Go to \lstinline|firesim-dosa/deploy|, and within \lstinline|config_hwdb.yaml| paste the contents of the file in \lstinline|built-hwdb-entries/| (there should be one file containing a YAML-formatted entry).

\subsection{Evaluation and expected results}

\subsubsection{Figure~\ref{fig:correlation}: Analytical model correlation with Timeloop}
On the \textbf{user machine}, run the following commands. This will correlate \sys{}'s differentiable model against Timeloop for our 10,000 point dataset and store the error plots to \lstinline|output_dir/error_<metric>.png|.
\begin{lstlisting}
cd dosa
./fig4.sh
\end{lstlisting}

\subsubsection{Figure~\ref{fig:search_curve}: Optimization of Gemmini-TL versus baseline algorithms}
In the same environment, run the following script, selecting one workload:
\begin{lstlisting}
./fig7.sh (unet|resnet50|bert|retinanet)
\end{lstlisting}

This will take several hours to run, per workload, and generate a plot at \lstinline|output_dir/network_searcher_<workload>_log_<timestamp>.png|. This corresponds to the plot to Figure 5, but over one run rather than averaged over 5. Results should fall within or close to the confidence bounds of the original plot.

\subsubsection{Figure~\ref{fig:arch_compare}: Comparison to hand-tuned accelerators}
Only after running \lstinline|fig7.sh| for the corresponding workload, run:
\begin{lstlisting}
./fig8.sh (unet|resnet50|bert|retinanet)
\end{lstlisting}
The plots will be generated at the location \lstinline|output_dir/arch_compare_<workload>_<timestamp>.png|. Since these are based on the results of one run rather than averaged over 5, results here will again vary slightly compared to the original plot.

\subsubsection{Figures~\ref{fig:gemmini-dnn} and \ref{fig:gemmini-dnn-not-training-set}: Gemmini-RTL performance prediction accuracy}
Run the following script:
\begin{lstlisting}
./fig10.sh
\end{lstlisting}

This will reproduce the plots in Figures~\ref{fig:gemmini-dnn} and \ref{fig:gemmini-dnn-not-training-set} under \lstinline|output_dir/predict_<predictor>_<dataset>.png|. These plots show the prediction accuracy of the three different predictors on the two datasets of Gemmini-RTL latency, which were previously generated using FireSim.

\subsubsection{Figure~\ref{fig:gemmini-results}: Optimization of Gemmini-RTL}
Now, \textbf{move to the AWS EC2 instance} set up with the FireSim fork. To run the full workflow of Figure~\ref{fig:gemmini-results}, we would need to train two DNN models, run \sys{} (constraining the number of PEs to 16x16), select the mappings with the best predicted performance, evaluate latency with FireSim, then combine with energy numbers from Accelergy. To reduce runtime and work that must be done across both the user machine and EC2 instance, we provide the mappings generated by \sys{} during this experiment directly to the evaluator as part of our FireSim fork. To build the software for a given workload and run FireSim, run the following:

\begin{lstlisting}
cd ~/firesim-dosa/target-design/chipyard/generators/gemmini/software/gemmini-rocc-tests
./artifact_script.sh (analytical|both|dnn) (unet|resnet50|bert|retinanet)
\end{lstlisting}

The first argument to \lstinline|artifact_script.sh| indicates which of the three latency predictors from the previous section should be used. The second argument indicates the target workload. This script launches FireSim automatically and should take a few minutes to run. Depending on the target workload, FireSim will generate either one or two directories under \lstinline|deploy/results-workload|, for matrix multiplication and/or convolutional layers. Pass the previously selected options, along with the directories (\lstinline|($result_dir_1)|, and potentially \lstinline|($result_dir_2)|) to the parsing script.

\begin{lstlisting}
cd ~/firesim-dosa/target-design/chipyard/generators/gemmini/software/gemmini-rocc-tests
python parse_results.py                        --pred (analytical|both|dnn)              --workload (unet|resnet50|bert|retinanet) --result ($result_dir_1)                   --result ($result_dir_2)
\end{lstlisting}

This will update the CSV file located at \lstinline|gemmini-rocc-tests/artifact/<predictor>/<workload>.csv|. \textbf{Copy this file back to the user machine}, to your choice of path \lstinline|($workload_csv)|. On the user machine, run the following to print out the EDP of the Gemmini default mapper/HW and the EDP of the mappings/HW found by \sys{}, all using latency numbers from FireSim. The relative magnitude of the Gemmini default and \sys{} EDPs should match those in Figure~\ref{fig:gemmini-results}.
\begin{lstlisting}
./fig12.sh (unet|resnet50|bert|retinanet) ($workload_csv)
\end{lstlisting}

When you are done evaluating, go to the EC2 console and terminate your instance(s).



\subsection{Methodology}

Submission, reviewing and badging methodology:

\begin{itemize}
  \item \url{https://www.acm.org/publications/policies/artifact-review-and-badging-current}
  \item \url{http://cTuning.org/ae/submission-20201122.html}
  \item \url{http://cTuning.org/ae/reviewing-20201122.html}
\end{itemize}

\end{document}